\documentclass{ieeeaccess}
\pdfoutput=1
\usepackage{cite}
\usepackage{amsmath,amssymb,amsfonts}
\usepackage{algorithmic}
\usepackage{graphicx}
\usepackage{textcomp}

\usepackage[hyphens]{url}
\usepackage{graphicx}
\usepackage{csquotes} 
\usepackage{xcolor} 
\usepackage{ragged2e}
\usepackage{subfigure}
\usepackage{amsthm}
\usepackage[per-mode = symbol]{siunitx}
\usepackage{verbatim}
\usepackage{array}
\usepackage{hyperref}

\newcommand{\RN}[1]{%
  \textup{\uppercase\expandafter{\romannumeral#1}}%
}

\usepackage{enumerate}
\usepackage[shortlabels]{enumitem}
\usepackage{microtype} 
\usepackage{adjustbox} 
\usepackage[all]{nowidow}
\usepackage{lipsum}
\usepackage{listings}
\usepackage{booktabs} 
\usepackage{etoolbox}

\usepackage{hologo}
\usepackage{listings}
\AtBeginEnvironment{quote}{\itshape}

\newcounter{requirement}[section]
\newenvironment{requirement}[1][]{\refstepcounter{requirement}\par\medskip
  \noindent \textbf{Requirement~\therequirement. #1} \rmfamily}{\medskip}

\lstdefinestyle{swrl}{
    basicstyle=\small\ttfamily,
    backgroundcolor=\color{gray!10}, 
    breaklines=true, 
    frame=single, 
    framesep=5pt, 
    frameround=tttt, 
    numbers=left, 
    numberstyle=\tiny\ttfamily\color{gray}, 
    language=Java, 
    showstringspaces=false, 
    tabsize=4, 
    captionpos=b, 
}

\def\BibTeX{{\rm B\kern-.05em{\sc i\kern-.025em b}\kern-.08em
    T\kern-.1667em\lower.7ex\hbox{E}\kern-.125emX}}
\begin{document}

\thispagestyle{empty}
\twocolumn[
\begin{@twocolumnfalse}
	%
	%
	
	\large {\copyright\ 2024 IEEE. Personal use of this material is permitted. Permission from IEEE must be obtained for all other uses, in any current or future media, including reprinting/republishing this material for advertising or promotional purposes, creating new collective works, for resale or redistribution to servers or lists, or reuse of any copyrighted component of this work in other works.} \\ \\
	
	{\Large Published in \emph{IEEE Access}} \\ \\
 
	{\Large DOI: \href{https://doi.org/10.1109/ACCESS.2024.3494036}{10.1109/ACCESS.2024.3494036}} \\ \\
	
	Cite as:
	\vspace{0.1cm}
	
	\noindent\fbox{%
		\begin{minipage}{0.98\textwidth}%
			N.~F.~Salem, M.~Nolte, V.~Haber, T.~Menzel, H.~Steege, R.~Graubohm, and M.~Maurer, ``{An} {Ontology-based} {Approach} {Toward} {Traceable} {Behavior} {Specifications} in {Automated} {Driving},'' \emph{IEEE Access}, vol. 12, pp. 165203-165226, 2024. DOI: 10.1109/ACCESS.2024.3494036.
		\end{minipage}
	}
	\vspace{2cm}
	
\end{@twocolumnfalse}
]

\noindent%
\hologo{BibTeX}:

\noindent
	\begin{centering}
	\footnotesize
	\begin{lstlisting}[frame=single,linewidth=\textwidth]
@article{salem_risk_2024,
    title = {{An} {Ontology-based} {Approach} {Toward} {Traceable} {Behavior} {Specifications} in {Automated} {Driving}},
    volume = {12},
    journal = {IEEE Access},
    author = {Salem, Nayel Fabian and Nolte, Marcus and Haber, Veronica and Menzel, Till and Steege, Hans and Graubohm, Robert and Maurer, Markus},
    year = {2024},
    pages = {165203-165226},
}
	\end{lstlisting}
\end{centering}


\history{Date of publication xxxx 00, 0000, date of current version xxxx 00, 0000.}
\doi{XX.1109/ACCESS.20XX.0122113}

\title{An Ontology-based Approach Towards Traceable Behavior Specifications in Automated Driving}
\author{
\uppercase{Nayel Fabian Salem}\authorrefmark{1},
\uppercase{Marcus Nolte}\authorrefmark{1},
\uppercase{Veronica Haber}\authorrefmark{2},
\uppercase{Till Menzel}\authorrefmark{1},
\uppercase{Hans Steege}\authorrefmark{3},
\uppercase{Robert Graubohm}\authorrefmark{1},
and \uppercase{Markus Maurer}\authorrefmark{1}
}

\address[1]{Institute of Control Engineering, Technische Universit\"at Braunschweig, Braunschweig, Germany (e-mail: \{n.salem, marnolte, t.menzel, robert.graubohm, markus.maurer\}@tu-bs.de)}
\address[2]{PROSTEP AG, Munich, Germany (e-mail: veronica.haber@prostep.com)}
\address[3]{Institute of Economics and Law, Universität Stuttgart, Stuttgart, Germany (e-mail: hans.steege@ivr.uni-stuttgart.de)}
\tfootnote{This work was supported by the German Federal Ministry for Economic Affairs and Climate Action within the projects ``Verifikations- und Validierungsmethoden automatisierter Fahrzeuge im urbanen Umfeld (VVMethods)'' and ``Automatisierter Transport zwischen Logistikzentren auf Schnellstraßen im Level 4 (ATLAS-L4)''.
VVMethods is a successor project to the project ``Projekt zur Etablierung von generell akzeptierten Gütekriterien, Werkzeugen und Methoden sowie
Szenarien und Situationen zur Freigabe hochautomatisierter Fahrfunktionen (PEGASUS)'' and a Project in the PEGASUS Family.
}

\markboth
{Salem \headeretal: An Ontology-based Approach Towards Traceable Behavior Specifications in Automated Driving}
{Salem \headeretal: An Ontology-based Approach Towards Traceable Behavior Specifications in Automated Driving}

\corresp{Corresponding author: Nayel Fabian Salem (e-mail: n.salem@tu-bs.de).}

\begin{abstract}
Automated vehicles in public traffic are subject to a number of expectations: Among other aspects, their behavior should be safe, conforming to the rules of the road and provide mobility to their users \cite{thornton_incorporating_2017, nolte_representing_2018}. This poses challenges for the developers of such systems: Developers are responsible for specifying this behavior, for example, in terms of requirements at system design time. As we will discuss in the article, this specification always involves the need for assumptions and trade-offs. As a result, insufficiencies in such a behavior specification can occur that can potentially lead to unsafe system behavior.
In order to support the identification of specification insufficiencies, requirements and respective assumptions need to be made explicit.
In this article, we propose the \emph{Semantic Norm Behavior Analysis} as an ontology-based approach to specify the behavior for an Automated Driving System equipped vehicle.
We use ontologies to formally represent specified behavior for a targeted operational environment, and to establish traceability between specified behavior and the addressed stakeholder needs.
Furthermore, we illustrate the application of the Semantic Norm Behavior Analysis 
in a German legal context
with two example scenarios and evaluate our results.
Our evaluation shows that the explicit documentation of assumptions in the behavior specification supports both the identification of specification insufficiencies and their treatment. Therefore, this article provides requirements, terminology and an according methodology to facilitate ontology-based behavior specifications in automated driving.
\end{abstract}


\titlepgskip=-21pt

\maketitle

\section{Introduction}
\label{sec:introduction}
\PARstart{W}{ith} the shift from assisted to automated driving, Automated Driving Systems (SAE Levels 3 and higher \cite{noauthor_taxonomy_2021}) take over full control of the dynamic driving task under specified conditions -- depending on the level of automation. While the responsibility for maneuvering through traffic lies with the human driver in a conventional vehicle (SAE Levels 2 and lower \cite{noauthor_taxonomy_2021}), this responsibility lies with the Automated Driving System in case of an automated vehicle.\footnote{We use the term \enquote{automated vehicle} for sake of brevity. In this article, the term refers to the same concept as an \enquote{Automated Driving System equipped vehicle} according to the SAE~J3016 taxonomy \cite[Sec. 3.32.2]{noauthor_taxonomy_2021}.} As a result, developers of Automated Driving Systems need to shift from assuring safe interaction between a driver assistance system and a driver to assuring safe behavior for the interaction between an automated vehicle and its surrounding traffic.

To that end, ISO~21448~\cite{noauthor_road_2022-1} specifies guidelines for the \emph{safety of the intended functionality}. This includes the evaluation of potentially hazardous behavior due to the behavior specified in a system's development \cite[Clause~10.1]{noauthor_road_2022-1}. However, the standard does not define \emph{how} behavior should be specified. 

A scenario-based approach is suggested in industry standards and reports \cite{noauthor_road_2022-1, noauthor_road_2023, noauthor_road_2020}, research works \cite{schuldt_effiziente_2013, bagschik_ontology_2018, menzel_scenarios_2018}, and regulatory documents \cite{noauthor_commission_2022} as a means to structure the operational environment of an automated vehicle. Ulbrich~\emph{et~al.} define a scenario as \enquote{the temporal development between several scenes in a sequence of scenes. Every scenario starts with an initial scene. Actions \& events as well as goals \& values may be specified to characterize this temporal development in a scenario. Other than a scene, a scenario spans a certain amount of time} \cite[p.~986]{ulbrich_defining_2015}. 
In this context, we use the concept of \emph{maneuvers} to describe the behavior of an agent at the transition between two subsequent scenes in a scenario. According to Jatzkowski~\emph{et al.} (translated by us) a maneuver is defined as: \enquote{a logical abstraction of possible vehicle trajectories. In an automata-theoretic sense, a maneuver represents a discrete state, which is only partially observable from an external perspective. Every maneuver is tied to an intention} \cite[p.~17]{jatzkowski_zum_2021}.  If there is more than one possible maneuver for an actor in a scene, maneuver options specify the transitions from one scene to multiple subsequent scenes. Thus, we define a \emph{behavior specification} as a consistent set of requirements that specify maneuver options, which an automated vehicle may perform under specified conditions. Scenarios provide a means to structure these operating conditions and can therefore support a behavior specification.

\Figure[h!](topskip=0pt, botskip=0pt, midskip=0pt)[width=0.9\columnwidth]{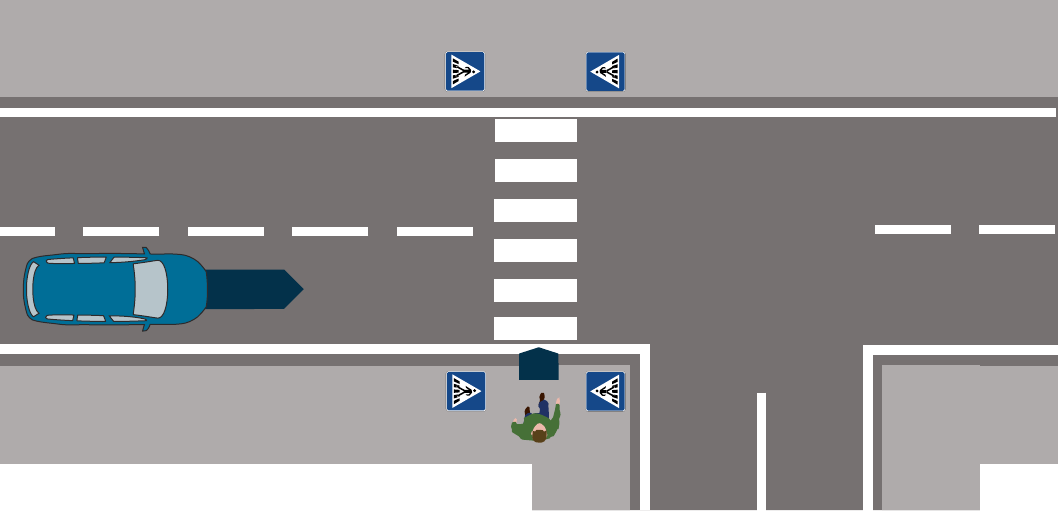}
{Example Scenario: A pedestrian attempts to use a pedestrian crossing. The ego-vehicle approaches the pedestrian crossing from the left. \label{fig:examp}}

An example scenario that we will use for illustration throughout this article is shown in \autoref{fig:examp}. In this scenario, an automated ego-vehicle is approaching a pedestrian crossing. At the same time, a pedestrian attempts to cross the road at the marked area of the pedestrian crossing. Intuitively, a behavior specification would specify a requirement that the ego-vehicle shall perform a stopping maneuver at the pedestrian crossing to enable the pedestrian to cross the road. However, in a development process, developers not only need to specify maneuvers the vehicle is supposed to perform, but they also need to specify the operating conditions (e.g. the entities populating the traffic context) under which the maneuver should be performed. For example, the blue traffic signs and the markings on the road (\autoref{fig:examp}) indicate a pedestrian crossing according to the German traffic code. Additionally, the operating conditions can be varied to cover other scenarios an automated vehicle can encounter. For example, developers could ask: \emph{Which maneuver is appropriate, if the pedestrian does not use the marked area of the pedestrian crossing to cross the road?} The answer to this question requires interpretation depending on the specific traffic context. Another determining factor for what is considered reasonable in such specifications is the legal context in which the automated vehicle is deployed. For example, it can be reasonable to assume that an adult would wait for the vehicle to pass until crossing the road. In other instances, a child could cross the road thinking that it has right of way even if it does not cross the road within the marked area of the pedestrian crossing. In each of these examples, it is necessary to make decisions and assumptions in the specification of expected behavior explicit, to identify scenarios in which the specified behavior potentially leads to harm.\footnote{ISO~21448 focuses on scenarios in which the risk of harm is assessed as
unreasonable. As risk assessment is not in scope of this article, we confine ourselves to instances, where harm can occur. Subsequently, some of these scenarios can be assessed to include a reasonable level of risk. In previous work, we elaborated on an approach to assess and treat risks in the context of a behavior specification \cite{salem_risk_2024}.}
Additionally, multiple stakeholders have potentially conflicting expectations that need to be consolidated in order to specify a consistent set of expected maneuvers. Passengers of the automated vehicle could expect the automated vehicle to only stop for pedestrians if there is an imminent collision. However, pedestrians could expect the automated vehicle to slow down early to respect their right of way.

For this scenario, we show that developers must decide on reasonable assumptions under specified conditions during the specification of behavior. 
In contrast to approaches that aim at a general algorithmic solution to these challenges, scenario-based approaches enable an explicit documentation of, for example, the considered operating conditions. 
To that end, ISO~21448 requires the identification of functional insufficiencies resulting from such assumptions and decisions in order to ensure the safety of the intended functionality. While ISO~21448 (implicitly) requires the identification of insufficiencies in the behavior specification of an automated vehicle, the standard does not provide guidance on how to satisfy this requirement. As decisions and assumptions, which are an inherent part of a behavior specification, can lead to specification insufficiencies, methods are needed to explicitly represent the specified behavior of an automated vehicle.

In the context of a scenario-based approach, the process of specifying the behavior of an automated vehicle involves mainly two steps.
First, developers need to elicit the expectations and needs of relevant stakeholders with respect to the behavior of an automated vehicle. However, these expectations and needs can be rather vague and ambiguous and have to be specified in formal stakeholder requirements.\footnote{Details in \autoref{sec:term}}
Second, stakeholder (needs and) requirements can be conflicting, especially if they are specified without a precise context. As a result, a consolidation of stakeholder requirements is necessary to obtain a sound behavior specification.

In this article, we introduce an ontology-based approach -- the \emph{Semantic Norm Behavior Analysis (SNBA)} -- to facilitate the explicit specification of behavior for automated vehicles. We use scenarios to describe operating conditions and specified behavior in terms of maneuver options that satisfy stakeholder requirements. In the context of a behavior specification, we will show how scenarios can be used to facilitate the specification and consolidation of expected behavior. Additionally, we formalize the specification in ontologies. These ontologies support the human-readable specification with machine-readable formal logic. The representation of the behavior specification in formal logic enables automated reasoning for consistency of the specification. 
We use ontologies to capture the specified behavior -- especially maneuver options -- and the sources that have been used to specify these maneuvers. Such sources are, for example, the traffic code that applies in the targeted operational environment. These sources represent a set of stakeholder needs that are analyzed throughout the behavior specification. With the proposed ontology-based approach, we seek to establish traceability between those needs and the resulting specified behavior.

In summary, the contributions that are presented in this article include:

\begin{itemize}
    \item terminology that supports the behavior specification task,
    \item a method for an explicit behavior specification of an automated vehicle, and
    \item representations of a behavior specification to support the communication between domain experts (e.g., regulators, safety engineers, and systems engineers).
\end{itemize}

To provide these contributions, we first elaborate on existing approaches in the literature in \autoref{sec:relw}. Based on our literature review, we elicit requirements for the behavior specification of automated vehicles (\autoref{sec:req}). Additionally, we propose terminology (\autoref{sec:term}) to describe key concepts in a behavior specification. Based on the formulated requirements, we introduce the Semantic Norm Behavior Analysis (\autoref{sec:snba}).
We present our results from a subsequent example application of the method using multiple representations in \autoref{sec:examp}. 
This example is limited to a German legal context, as this scope lies within our field of expertise.
Finally, we discuss contributions and limitations of the proposed method with respect to our elicited requirements (\autoref{sec:eval}).

\section{Related Work}
\label{sec:relw}
In this section, we review related work in order to motivate the importance of a behavior specification and to highlight open questions based on existing approaches on specifying behavior of automated vehicles.

Several standards and technical reports related to automated driving emphasize, at least implicitly, the importance of formalized behavior specifications. 

ISO~21448~\cite{noauthor_road_2022-1} requires (among others) to identify insufficiencies of specification that lead to hazardous behavior and cause unreasonable risk. Such insufficiencies can arise in the specification of the intended functionality at the vehicle-level. A vehicle-level specification can include the specification of intended behavior. In this context, ISO~21448 introduces the \emph{vehicle-level SOTIF strategy}. The vehicle-level SOTIF strategy is defined as a \enquote{set of vehicle-level requirements for the intended functionality used to support design, verification and validation activities to achieve the SOTIF}~\cite[Clause~3.34]{noauthor_road_2022-1}. This definition allows for the specification of intended behavior as part of the vehicle-level SOTIF strategy. 
In combination with ISO~21448 requirements on traceability, a formalized process for the specification of intended behavior is required to support SOTIF activities. 
However, ISO~21448 does not provide guidance on how to specify vehicle-level requirements for the intended behavior.

ISO/TR~4804~\cite{noauthor_road_2020} states that the safety of an Automated Driving System can be addressed by \enquote{designing scenario-based system behaviours}~\cite[p.~19]{noauthor_road_2020}. 
To that end, ISO/TR~4804 follows ISO~21448 (ISO/PAS 21448 at the time) and highlights the importance of analytical approaches to ensure system safety.
An application of scenario-based system behaviors for the identification of hazardous events was, for example, shown in \cite{graubohm_towards_2020}. Additionally, the authors of the ISO/TR~4804 highlight challenges, especially regarding the role of traffic rules. While some traffic rules may be directly applicable to multiple driving contexts (such as the meaning of a STOP-sign) other rules could require interpretation by the developers \cite[Clause~5.3.2.10]{noauthor_road_2020}. As a result, there is a need for methods that must also support the documentation of scenario-based interpretations of traffic rules.

IEEE~Std~2846-2022~\cite{noauthor_ieee_2022} provides a minimum set of assumptions that shall be considered by safety-related models. 
The standard highlights the importance of explicit assumptions in the design of safety-related models for automated vehicles and applies these assumptions at a behavioral level rather than focusing on specific functions. However, IEEE~Std~2846-2022 is explicitly limited to a minimum set of kinematics-based assumptions and mentions the need for further refinement by regulators or manufacturers. For example, it is left open how the specification of an operational design domain and the intended behavior of an automated vehicle influence the necessary assumptions that are described in IEEE~Std~2846-2022.

From an academic research perspective, there are multiple publications related to the behavior specification of automated vehicles. 

Lacking the availability of ISO~21448, Reschka~\cite{reschka_fertigkeiten-_2017} describes a process to compile an item definition for an automated vehicle based on requirements from ISO~26262~\cite{noauthor_road_2018}. In his proposal, Reschka~\cite{reschka_fertigkeiten-_2017} mentions the importance of explicitly specifying behavior to identify necessary capabilities of the automated vehicle. While Reschka provides initial guidelines, the described specification process requires further formalization, for example to identify inconsistencies of a behavior specification in a set of scenarios.

Meyer~\emph{et~al.}~\cite{meyer_scenario-_2022} focus on utilizing techniques from the domain of model-based systems engineering (MBSE) for a SOTIF-compliant design of automated driving functions. In their work, Meyer~\emph{et~al.}~\cite{meyer_scenario-_2022} propose to first model the expected behavior in the operational environment before the derivation of sub-system requirements. In order to model expected behavior, the authors follow a scenario-based development paradigm. Meyer~\emph{et~al.} claim full traceability between scenarios and system requirements. However, the causal relations between scenario-specific context elements to internal and external behavior are not explicitly described. Additionally, the expressiveness of the maneuver concept used by Meyer~\emph{et~al.} depends on the respective block definition diagrams that are mentioned but not presented. Hence, the expressiveness of the maneuver concept in \cite{meyer_scenario-_2022} is limited.

To support the explicitness and consistency of behavior specifications, Irvine~\emph{et~al.}~\cite{irvine_structured_2023} propose an intermediate representation in the specification process. The authors present a structured language for expressing the rules of the road. While such a semi-formal natural language facilitates traceability of concepts to their natural language sources (i.e., stakeholder needs), the behavior specification can be further supported by formal machine-readable representations \cite{irvine_structured_2023}. 

In the context of scenario generation, Bagschik~\emph{et~al.}~\cite{bagschik_ontology_2018, bagschik_wissensbasierte_2018, bagschik_systematischer_2022} present a knowledge-based approach. The approach uses ontologies to create sequences of scenes (i.e. scenarios) based on possible maneuver options of the modeled agents. As a result, some of the generated scenarios can be discarded as part of system design and development. For example, there are maneuver options of an ego-vehicle that are logically possible but would be a deliberate violation of traffic rules. To that end, Da Costa~\emph{et~al.}~\cite{da_costa_ontology-based_2024} propose an ontology-based approach to generate (test) scenarios based on formal rules in order to constrain the scenario generation task. As the formal rules are based on legal sources, Da Costa~\emph{et~al.} point out that interpretation of such sources needs a formalized approach.

Ontologies are a common tool for formalizing required domain knowledge.
They are often used, for example, as a foundation for expert systems \cite{mora_development_2022} and have also been applied to the problem of runtime decision-making for decades in the robotics domain \cite{manzoor_ontology-based_2021}. 
Also, the field of automated driving has seen a fair share of ontology-based approaches for situation representation \cite{hulsen_traffic_2011, buechel_ontology-based_2017, armand_situation_2016, armand_ontology-based_2014} and decision-making \cite{zhao_fast_2016}, respectively.
Our approach builds on similar ideas, regarding the formalization of knowledge about the operational environment.
A key difference to runtime applications is our focus on the development process: We consider a behavior specification as a \emph{design-time} artifact which is created during design and development in order to support requirements formulation and to provide a basis for system architecture(s) and the technical implementation of an automated vehicle. As the actual operating conditions, the system will need to operate in, can never be fully defined at design time, developers need, amongst others, to make assumptions regarding foreseeable operating conditions and specify desired driving maneuvers. 
The resulting methodology hence has a slightly different scope: We put emphasis on traceable knowledge formalization and use ontologies as a tool to document such assumptions as a means for traceability.

The formalization of behavioral rules was addressed by related work as well.
Rizaldi~\emph{et~al.}~\cite{rizaldi_formalising_2015, rizaldi_formalising_2017}, Nikol~\emph{et~al.}~\cite{nikol_formalisierung_2019}, Maierhofer~\emph{et~al.}~\cite{maierhofer_formalization_2020}, and Westhofen~\emph{et~al.}~\cite{westhofen_towards_2022} propose a direct translation of the German Road Traffic Regulation into formal logic (i.e., higher-order logic and temporal logic). These approaches apply techniques from computational law to the domain of automated driving. Genesereth defines that \enquote{Computational Law is the branch of legal informatics concerned with the automation of legal reasoning.}\cite{genesereth_what_2021}
However, the validity of such approaches can be questioned from a legal perspective (cf. \cite{gstottner_durfen_2021, steege_automatisierte_2022} for legal perspectives written in German). 

From a safety engineering perspective, such approaches can be criticized as well. The direct translation of (intentionally vague) natural language traffic rules into mathematical expressions involves both qualitative and quantitative assumptions. These assumptions are implicitly encoded in the resulting formal specification and are highly dependent on the considered scenarios. Therefore, a directly generated formal rule catalog lacks traceability to the underlying assumptions in the specification. 
Such a missing traceability of the specification is especially challenging in the context of showing compliance with ISO~21448~\cite{noauthor_road_2022-1} as traceability of the specification and design is explicitly required by the standard \cite[Clause~5.3]{noauthor_road_2022-1}.

To account for the challenge of full compliance with a formal set of rules in an open context, Censi~\emph{et~al.}~\cite{censi_liability_2019} propose the \emph{rulebooks} formalism. Such pre-ordered sets of rules are supposed to include behavioral needs (legal, ethical, and cultural) in a behavior specification. Each rule is assigned with a violation metric which is used to assess compliance of a planned behavior with the rules. Finally, the planned behavior that is the least conflicting with the violation metrics and the specified hierarchy of rules is considered \enquote{optimal} with respect to the rulebook. While Censi~\emph{et~al.}~\cite{censi_liability_2019} explicitly exclude the runtime aspect of efficient motion planning with rulebooks, the authors do not mention how a rulebook can be selected depending on the situation. This leads to the question, whether one rulebook can be formulated that generalizes for a set of scenarios. This challenge is subsequently addressed in \cite{bogdoll_informed_2024} with an introduction of a situation-aware rulebook. However, both, the definition of any rulebook, and the situation-dependent selection of a rulebook depend on context-specific assumptions on appropriate rule hierarchies. As a result, rulebooks account for hierarchies in the behavior specification at runtime, but require explicit documentation at design time.

In scenario-based approaches, the performance of an automated vehicle is assessed based on its behavior in a set of scenarios. In order to formally describe behavior at an implementation-agnostic level, the Phenomenon-Signal Model (PSM) is introduced by Beck~\emph{et~al.}~\cite{beck_phenomenon-signal_2022, beck_phanomen-signal-modell_2021}. It captures causal relations from operating conditions in a scenario to specified \emph{facts}. Beck~\emph{et~al.}~\cite{beck_phenomenon-signal_2022, beck_phanomen-signal-modell_2021} propose a formalism that enables the inference of such causal relations based on stimuli from a scene (as defined by Ulbrich~\emph{et~al.}~\cite{ulbrich_defining_2015}) and a set of scenario-specific \emph{rules}. However, the Phenomenon-Signal Model does not provide guidelines on how to specify facts and rules in a traceable way, based on stakeholder needs that finally lead to the specified maneuver options. 

In this section, we reviewed literature related to the task of generating a behavior specification for an automated vehicle. These publications either highlight the importance of specifying behavior, provide general guidelines, or raise implicit requirements for approaches to compile a behavior specification. 
As a result, we propose an approach for behavior specification which uses ontologies to establish a traceable formalization of concepts that can be found, for example, in work from Beck~\emph{et~al.}~\cite{beck_phenomenon-signal_2022, beck_phanomen-signal-modell_2021} , Rizaldi~\emph{et~al.}~\cite{rizaldi_formalising_2015, rizaldi_formalising_2017}, and Irvine~\emph{et~al.}~\cite{irvine_structured_2023} and combine it with a model of the operational environment that is comparable to models by Armand~\emph{et~al.}~\cite{armand_situation_2016, armand_ontology-based_2014}, Bagschik~\emph{et~al.}~\cite{bagschik_ontology_2018, bagschik_wissensbasierte_2018, bagschik_systematischer_2022}, and Czarnecki~\cite{czarnecki_operational_2018, czarnecki_operational_2018-1}.
Ontology-based approaches come with their own set of challenges (e.g. with respect to scalability). We will elaborate on these limitations throughout this article and provide a dedicated discussion of limitations in \autoref{subsec:crit}.

\section{Requirements Elicitation}
\label{sec:req}
Based on our review of related work, we derive requirements for a behavior specification of an automated vehicle. More specifically, the elicitation of requirements focuses on the research gap of how a behavior specification can explicitly include context-specific assumptions and decisions.
The requirements also provide the basis for our evaluation of the Semantic Norm Behavior Analysis in \autoref{sec:eval}.
As we focus on the aspect of traceability in this article, these requirements are not exhaustive to cover the entire task of creating a behavior specification for an automated vehicle.

Censi~\emph{et~al.} describe the challenge of creating a behavior specification as \enquote{defining what the car is supposed to do}~\cite[p.~8536]{censi_liability_2019}. However, there are different perspectives from multiple stakeholders on what should be considered for the definition of compliant behavior. Hence, we deem elicited stakeholder needs as a key source of a behavior specification.

\begin{requirement}
    A behavior specification shall be explicitly derived from the needs of relevant stakeholders. \label{req:need}
\end{requirement}

In order to integrate the behavior specification in a development process for an automated vehicle, the specification needs to account for adaptions of the identified needs and specified requirements. Censi~\emph{et~al.}~\cite{censi_liability_2019} argue that, additionally, implementation limitations need to be considered for a specification of the intended behavior.

\begin{requirement}
    A behavior specification shall be adaptable to shifts in stakeholder needs and implementation limitations. \label{req:adapt}
\end{requirement}

While a behavior specification shall account for the needs of relevant stakeholders, these needs can be expected to include conflicts and may require interpretations depending on the context. As a result, maintaining consistency within the specification becomes a major challenge, but also a requirement (also formulated by Butz~\emph{et~al.}~\cite{butz_soca_2020}).

\begin{requirement}
    A behavior specification shall be a consistent representation of stakeholder requirements regarding the intended behavior in all considered scenarios. \label{req:consist}
\end{requirement}

Behavior is specified with respect to a set of operating conditions. In order to generate a consistent behavior specification, the description of these operating conditions needs to be consistent as well.
Glatzki~\emph{et~al.} define the concept of \emph{atomic behavior spaces} as \enquote{a segment of the road network within which the behavioral attributes do not change}~\cite[p.~670]{glatzki_behavioral_2021}. Similarly, Butz~\emph{et~al.}~\cite{butz_soca_2020} describe \emph{zone graphs} as an abstraction of the scenery. Both approaches model the operational environment such that behavioral constraints \cite{glatzki_behavioral_2021} or even the intended behavior \cite{butz_soca_2020} can be specified on a consistent basis.

\begin{requirement}
    A behavior specification shall refer to a consistent representation of the operational environment. \label{req:operat}
\end{requirement}

Furthermore, Glatzki~\emph{et~al.}~\cite{glatzki_behavioral_2021, glatzki_inferenz_2022} formulate \emph{behavioral attributes} that must be included in a scenery description. To reduce the complexity of the behavioral requirements, they rely on the scenery as a commonalty between multiple scenarios. However, the authors, for example, do not include the behavior of other traffic agents in their approach. We understand that the reason to specify behavioral attributes in a scenery is to capture behavioral constraints that apply across multiple scenarios. As a result, we formulate a more general requirement for a behavior specification.

\begin{requirement}
    A behavior specification shall only contain information from the scenario description that influences the behavioral requirements. \label{req:scen}
\end{requirement}

Reschka~\cite{reschka_fertigkeiten-_2017} outlines a possible utilization of the behavior specification for the identification of necessary \emph{capabilities} in the development process of an automated vehicle. A comparable concept is introduced by Nowakowski~\emph{et~al.}~\cite{nowakowski_development_2015, nowakowski_determining_2016} with \emph{behavioral competency}. \enquote{Behavioral competency refers to the ability of an AV [(automated vehicle)] to operate in the traffic conditions that it will regularly encounter [...]}~\cite[p.~141]{nowakowski_development_2015} The automotive industry has adopted the concepts of behavioral competency \cite{waymo_llc_waymo_2021, noauthor_avsc_2021} and capability \cite{noauthor_road_2020}, especially as a way to implement safety-by-design practices. While some capabilities focus on, for example, technical aspects such as \emph{detect[ing] when degradation is not available}~\cite{noauthor_road_2020}, other capabilities are directly linked to the behavior of an automated vehicle, such as \emph{creat[ing] a collision-free and lawful driving plan}. Thus, consistency of specified capabilities that presume certain behaviors and the behavior specification itself needs to be maintained.

\begin{requirement}
    A behavior specification shall facilitate the identification of necessary capabilities (or behavioral competencies) of an Automated Driving System. \label{req:capab}
\end{requirement}

In addition to the specification of intended behavior itself, the transition to the design of the technical systems that compose the automated vehicle also needs to be considered.
ISO~21448~\cite{noauthor_road_2022-1} introduces the vehicle-level SOTIF strategy that should specify the driving policy. Such a specification can \enquote{influence the design of all the building blocks of an ADS-equipped vehicle} \cite[Clause~D.1.1]{noauthor_road_2022-1}. Therefore, a vehicle-level SOTIF strategy needs to specify a driving policy that does not presume a specific technical solution for the implementation of the specified behavior. 

\begin{requirement}
    A behavior specification shall specify behavior independently of the technical solution. \label{req:tech}
\end{requirement}

In the \emph{rulebooks} approach, Censi~\emph{et~al.}~\cite{censi_liability_2019} elaborate that a goal of their behavior specification is to assess different results from a behavior planner with respect to a predefined set of criteria (i.e. rules). 
Therefore, a behavior specification should not predefine a single compliant behavior, but instead should facilitate the comparison of multiple behaviors. 
In addition to this application in runtime monitoring, we identified a need for a behavior specification to support SOTIF activities. As a result, we derive a more general requirement as some behavior (both generated by a planner or observed, e.g. in a test case) needs to be evaluated with respect to its compliance with the behavior specification \cite[Clause~10.1]{noauthor_road_2022-1}. 

\begin{requirement}
    A behavior specification shall provide criteria and requirements for the assessment of multiple maneuver options. \label{req:assess}
\end{requirement}

Within a behavior specification for automated driving, stakeholder requirements formulate the intended behavior of an automated vehicle under a set of operating conditions. 
The intended behavior can be specified at different levels of abstraction.
As a result, some approaches utilize abstract behavioral descriptions at runtime \cite{arens_representation_2002, arens_behavioral_2003}, others highlight the importance of an abstract behavior specification for the design of automated vehicles \cite{butz_soca_2020, lippert_behavior-semantic_2022}. While a high level of abstraction is beneficial in early phases of the safety lifecycle to support SOTIF activities, some activities (e.g. testing specific sensor effects) require a more detailed description of scenarios \cite{menzel_scenarios_2018} including the respective behaviors. As a result, the level of abstraction in a behavior specification is determined based on a trade-off between handling complexity and providing implementation details.

\begin{requirement}
    A behavior specification shall support a specification of intended behavior on multiple levels of abstraction. \label{req:abstr}
\end{requirement}

Finally, this work is inspired by the Phenomenon-Signal Model. Hence, we define a constraint for our ontology-based approach towards behavior specification. Beck~\emph{et~al.}~\cite{beck_phenomenon-signal_2022} introduce the concepts of \emph{facts} and \emph{rules} to formally represent maneuver options based on stimuli from the operational environment. In our literature review, we identified facts and rules as a means to support the traceability of a behavior specification from representations in natural language to machine-readable formats. Therefore, facts and rules are important concepts in our proposed ontology.

\section{Terminology}
\label{sec:term}
In this section, we compile the terms and definitions that build the foundation of our contribution. 

Since our focus is the specification of behavior for automated driving, we need to define the term \emph{behavior}. However, there are multiple definitions of behavior. According to Avi\u zienis~\emph{et~al.} behavior is defined as \enquote{what the system does [...] and described by a sequence of states}~\cite[p.~12]{avizienis_basic_2004}.
Additionally, Arkin defines behavior from the perspective of behavior-based robotics: \enquote{Behavior, simply put, is a reaction to a stimulus}~\cite[p.~66]{arkin_behavior-based_1998}.
These definitions have in common that they both refer to the behavior that is externally observable. Arkin extends this understanding by also considering the stimuli from the operational environment that lead to a specific response as part of a system's behavior. 
Thus, in the context of safety engineering, Nolte~\emph{et~al.}~\cite{nolte_towards_2017} elaborate that a nuanced distinction between internal and external behavior (cf. \cite{avizienis_basic_2004}) is necessary.
We adopt this way of thinking about behavior in order to support the explicit specification of interactions between an automated vehicle and its environment. 

As our approach follows the scenario-based development paradigm, there is a strong dependency on the term \emph{scenario}. Ulbrich~\emph{et~al.} define a scenario as \enquote{[...] the temporal development between several scenes in a sequence of scenes. Every scenario starts with an initial scene. Actions \& events as well as goals \& values may be specified to characterize this temporal development in a scenario. Other than a scene, a scenario spans a certain amount of time}~\cite[p.~986]{ulbrich_defining_2015}. Additionally, a scenario-based development paradigm enables engineers to refine the behavior specification on different levels of abstraction. Menzel~\emph{et~al.}~\cite{menzel_scenarios_2018} propose \emph{functional}, \emph{logical} and \emph{concrete} scenarios to account for increasing maturity of the development process from the concept phase to verification and validation activities. Similarly, we consider the behavior specification to have multiple degrees of abstraction \cite{irvine_structured_2023}.

In the context of automated driving, the concepts \emph{target operational domain} and \emph{operational design domain} are important to capture operating conditions of an automated vehicle. However, both concepts focus on different aspects. According to ISO~34503, the target operational domain is composed by the \enquote{set of operating conditions in which an ADS will be expected to operate}~\cite[Clause~3.7]{noauthor_road_2023}. In contrast, the operational design domain is defined as \enquote{operating conditions under which a given driving automation system or feature thereof is specifically designed to function [...]}\cite[Clause~3.21]{noauthor_taxonomy_2021}. Therefore, the specification of an operational design domain requires details from system design whereas the target operational domain focuses on the context an automated vehicle is deployed in. Since we already specified the importance of solution-independence of a behavior specification in Requirement \ref{req:tech}, the term target operational domain provides an important reference.

The terms \emph{stakeholder needs} and \emph{stakeholder requirements} are frequently used in the systems engineering standard ISO/IEC/IEEE~15288~\cite{noauthor_systems_2015}. However, no formal definition is provided. 
Faisandier~\emph{et~al.}~\cite{faisandier_stakeholder_2023} describe the goal of \emph{stakeholder needs and requirements definition activities} as the elicitation of clear and concise needs, which is followed by a transformation into consolidated and verifiable stakeholder requirements.
Stakeholder needs may contain vague and ambiguous statements. In contrast, stakeholder requirements are more \enquote{engineering-oriented}~\cite{faisandier_stakeholder_2023} (i.e. formulated in requirements-specific language) to enable a proper system definition. We adopt this description of stakeholder needs and requirements. In the context of behavior specification, we specify behavior with a consistent set of stakeholder requirements. We refer to such requirements as \emph{behavior-related stakeholder requirements} or \emph{behavioral requirements}.

Finally, the term \emph{ontology} needs to be specified. The term \emph{ontology} has different definitions depending on the application domain. We refer to ontologies as defined by Gruber~\cite{gruber_translation_1993} in the domain of computer science. \enquote{A common ontology defines the vocabulary with which queries and assertions are exchanged among agents} \cite[p.~201]{gruber_translation_1993}. Therefore, we focus on the communicative aspect of an ontology as a common and formally expressed set of terms in a knowledge base.

Given the introduction of foundational terms we introduce our proposed terminology. 
To account for the complexity of an automated vehicle's operational environment and for the heterogeneity of stakeholder needs we define the term \emph{norm behavior}:

\begin{quote}
    Specification of the (internal and external) behavior an actor shall exhibit in a scenario-overarching context resulting from legal, societal, and ethical needs.
\end{quote}

In \autoref{fig:def} norm behavior is defined as a subset of stakeholder needs with two main characteristics. First, norm behavior may include conflicting needs towards a vehicle's behavior that require consolidation. Second, norm behavior may be specified without referring to a consistent set of operating conditions that are described in a scenario (i.e. norm behavior is scenario-overarching).

To satisfy the requirements, which we elicited in \autoref{sec:req}, the concept of norm behavior is necessary but not sufficient. The documentation of assumptions and decisions that are necessary to design an automated vehicle requires the consolidation (and interpretation) of potentially conflicting stakeholder needs. In the context of a behavior specification, we propose to use scenarios as a common ground for such a consolidation process. We introduce the term \emph{target behavior} to capture consolidated stakeholder requirements with respect to the behavior of an automated vehicle:

\begin{quote}
    Specification of the (internal and external) behavior an actor shall exhibit in a scenario-specific context based on the specified norm behavior.
\end{quote}

According to Faisandier~\emph{et~al.}~\cite{faisandier_stakeholder_2023} stakeholder requirements can be characterized by properties of clarity and conciseness. Hence, target behavior is defined as a subset of stakeholder requirements in \autoref{fig:def} and is characterized by being consolidated and scenario-specific.

\Figure[ht](topskip=0pt, botskip=0pt, midskip=0pt)[width=0.9\columnwidth]{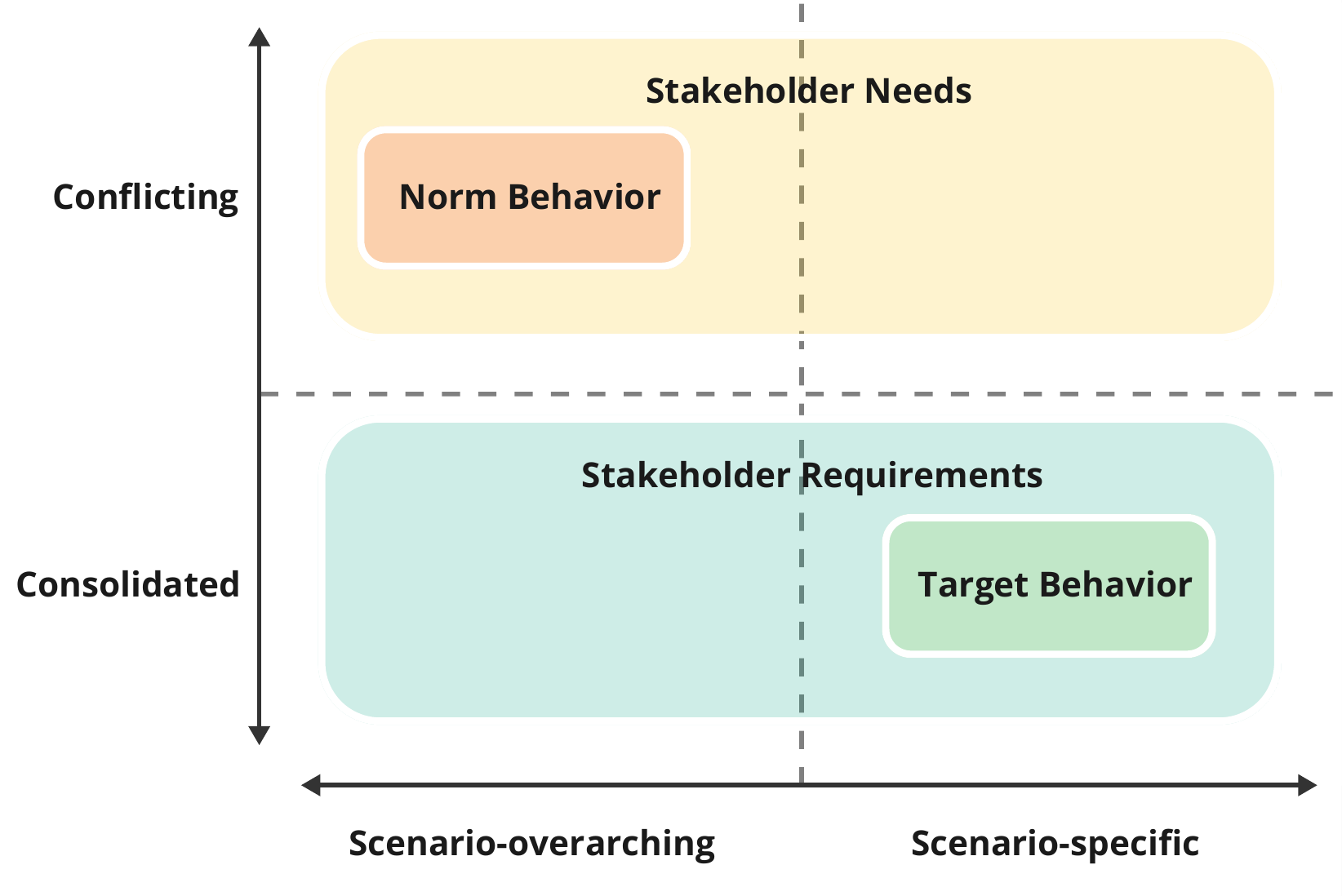}
{Key concepts \textit{target behavior} and \textit{norm behavior} introduced in this article and their relation to stakeholder needs and stakeholder requirements. \label{fig:def}}

Based on our definitions of \emph{norm behavior} and \emph{target behavior}, we will elaborate on their relations between each other and on the context they are defined in. As norm behavior specifies stakeholder needs, it is important to note that these needs can have different priorities and can even be conflicting. 
Therefore, we propose to separate the process of eliciting stakeholder needs in a scenario-overarching context (i.e., the target operational domain) and finally resolving conflicts in a scenario-specific context. This separation 
facilitates an explicit documentation of assumptions in the specification of target behavior as the behavioral requirements need to be specified under specific operating conditions. 

The terminology we described in this section sets the scope of this work. Moreover, the introduction of the terms \emph{norm behavior} and \emph{target behavior} provides the basis for the Semantic Norm Behavior Analysis. 

\Figure[h!](topskip=0pt, botskip=0pt, midskip=0pt)[width=1.9\columnwidth]{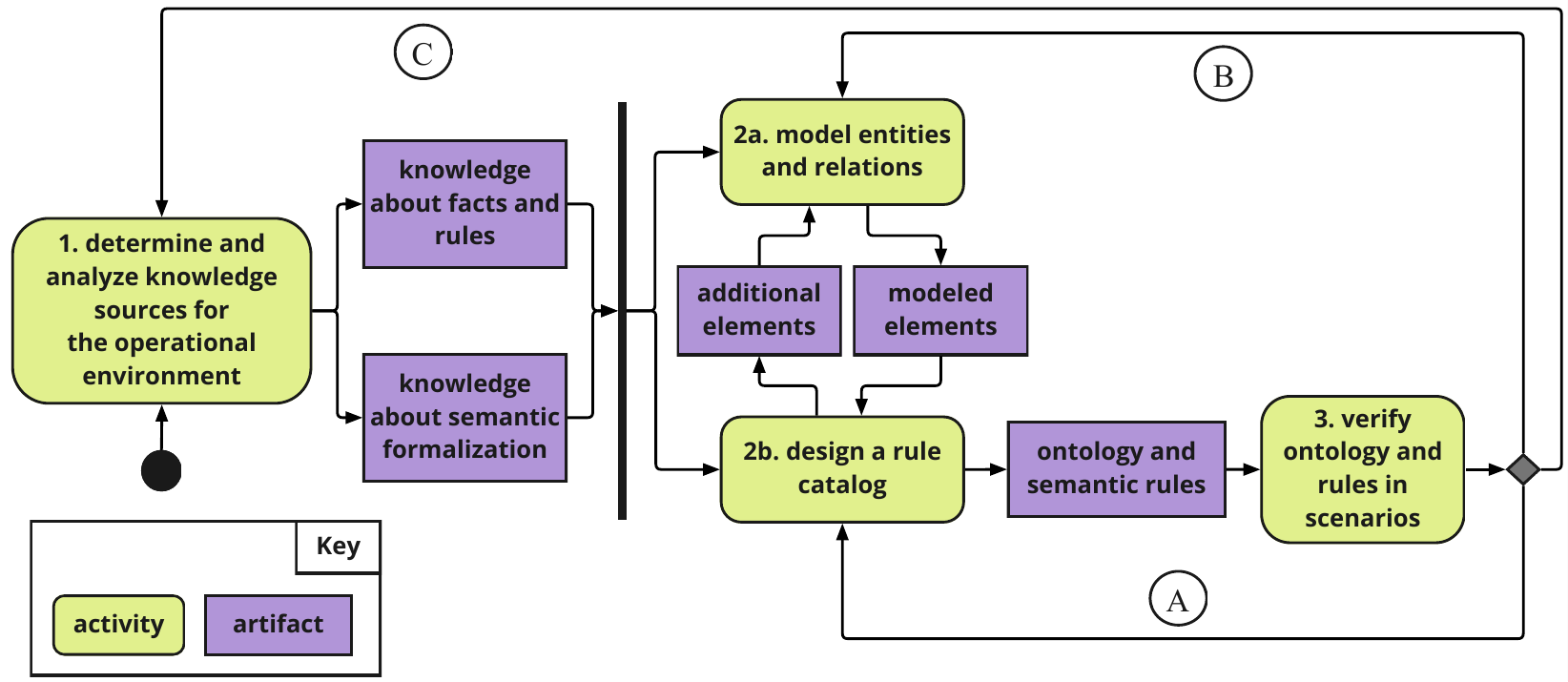}
{Schematic description of the Semantic Norm Behavior Analysis containing four basic activities. In activity 1 knowledge sources are determined and analyzed. Based on the resulting knowledge, facts and relations are modeled in an ontology (activity 2a). Subsequently rules are defined to infer maneuver options based on facts in a start scene (activity 2b). Finally, the consistency of the ontology and the rules is verified by using a logic reasoner. Modeling errors are accounted for by the respective iteration loop (\textit{A} -- rule errors; \textit{B} -- ontological errors; \textit{C} -- missing/false analysis of knowledge sources). \label{fig:snba}}

\section{Ontology-based Approach towards Behavior Specification}
\label{sec:snba}
In the previous sections, we argued that an explicit behavior specification can 
significantly contribute to the development of automated vehicles -- especially in the context of SOTIF.
First, we outlined the challenge for developers to address conflicting stakeholder needs with respect to a system's behavior. Second, we argued that ISO~21448 implicitly requires a specification of behavior to identify specification insufficiencies and to support the vehicle-level SOTIF strategy.

In this section we introduce the Semantic Norm Behavior Analysis. 
The method provides an ontology-based and thus formalized approach to specify target behavior.
Figure \ref{fig:snba} shows the key activities and artifacts of the Semantic Norm Behavior Analysis. We elaborate on each activity throughout this section. In \autoref{subsec:overv} we give an overview of the activities and artifacts that are part of the Semantic Norm Behavior Analysis. Subsequently, we specify these activities in more detail in \autoref{subsec:domainan}, \ref{subsec:ontol}, and \ref{subsec:vv}.

\subsection{Overview of the Semantic Norm Behavior Analysis}
\label{subsec:overv}
The Semantic Norm Behavior Analysis is designed to explicitly consider behavior-related stakeholder needs during behavior specification. These needs are often captured in laws, ethical guidelines or other documents. We refer to such documents as \emph{knowledge sources} for norm behavior.
At first, knowledge sources need to be defined and analyzed (\emph{Step~1} in \autoref{fig:snba}). These sources can contain both, knowledge about norm behavior and knowledge about domain-specific methods for the interpretation of norm behavior. For example, we will consider the German Road Traffic Regulation as a source of stakeholder needs for norm behavior. As the traffic code contains ambiguous statements, we use knowledge about legal analyses to give an example interpretation of the traffic code. This step of the Semantic Norm Behavior Analysis builds the foundation of the behavior specification as it supports the elicitation of stakeholder needs regarding the behavior of an automated vehicle. 
These elicited stakeholder needs can (and will) be ambiguous and conflicting. Following Systems Engineering practices, these needs must hence first be transformed into unambiguously formulated stakeholder requirements to provide a basis for the formulation of functional and technical requirements. This is a step that requires knowledge of skilled Systems- and Requirements Engineers who must make assumptions and trade-offs to come from stakeholder needs and stakeholder requirements to a functional and technical design.

In \emph{Step~2a} (\autoref{fig:snba}) the specified behavioral requirements are formalized in ontologies. 
We use ontologies (and rules) to assist this process: They provide the possibility to explicitly document required assumptions in the translation process (as we will show with the assumed crossing intention of a pedestrian in our example in \autoref{sec:examp}).
First and foremost, one ontology contains the operating conditions for which maneuver options need to be specified. Relevant entities are, for example, lanes, lane markings, and traffic signs. A second ontology formalizes expert knowledge in the behavioral requirements. Mainly, this ontology contains the causalities between a set of operating conditions and the specified maneuver options. Both ontologies provide a consistent description logic model that specifies the intended behavior in a scenario. 

As the reasoning capabilities of a description logic model are limited, we use logic programming rules to supplement the ontologies (\emph{Step~2b} in \autoref{fig:snba}).
Both activities (modeling entities and relations, and the formalization of rules) build an iterative process. 

\emph{Step~3} (\autoref{fig:snba}) includes the verification of the specified behavior. In this context, the ontologies and rules build a set of formal axioms that are evaluated regarding their consistency. The set of axioms is consistent, if description logic and logic programming rules can be applied to the ontologies without logical errors.
However, a proof of logical consistency is not sufficient to claim validity of the specified behavior. A validation step is required to show that the specified behavior complies with the considered stakeholder needs. As stakeholder needs can be formalized only partially, this validation step requires an expert-based evaluation. These experts, for example, need to assess legal compliance of a maneuver, if the traffic code is defined as one of the knowledge sources of stakeholder needs.

Finally, there are two key results that are generated with an application of the Semantic Norm Behavior Analysis. First, an ontology is generated that formally specifies target behavior for a set of scenarios. Second, the Semantic Norm Behavior Analysis requires the documentation of the rationale for why and how the specified behavior satisfies the considered stakeholder needs. In summary, such an ontology-based behavior specification contains both norm behavior and target behavior.

\subsection{Analysis of Norm Behavior from a Legal Perspective}
\label{subsec:domainan}
In \emph{Step~1} (\autoref{fig:snba}) of the Semantic Norm Behavior Analysis, knowledge sources are determined that represent relevant stakeholder needs.
These needs are formulated from multiple perspectives. However, 
an analysis of stakeholder needs highly depends on the domain of the experts that conduct the analysis. For example, a behavior specification based on laws and regulations requires different knowledge than a specification based on local driving styles. 
As the development of automated vehicles is of great legal importance, we focus on the application of our method from a legal perspective in this article.

In Germany, the Road Traffic Act (StVG\footnote{German abbreviation for \emph{Straßenverkehrsgesetz}}) and the Road
Traffic Regulation (StVO\footnote{German abbreviation for \emph{Straßenverkehrsordnung}}) provide important examples of normative legal sources for a behavior specification. Compliance with these documents is necessary in Germany, as the granting of type approval (in the European Union) depends on compliance of an automated vehicle's behavior with the local traffic rules. This also has direct liability implications for the car manufacturer because both, the German producer liability in accordance with Section 823 (1) of the German Civil Code (BGB), and the harmonized European Product Liability Directive, which was implemented in Germany in the Product Liability Act, refer to the design defect. Such a defect is deemed to exist if an average user cannot use the vehicle safely in accordance with its intended purpose. This justified safety expectations must be taken into account \cite{steege_haftung_2023}.
However, compliance with these legal documents is not straightforward as they require interpretation and the additional consideration of, for example, legal cases. Traceability of the behavior specification thus also takes on considerable importance in the context of product liability.

To that end, we use the German Road Traffic Act and the German Road
Traffic Regulation (s. \emph{knowledge about facts and rules} in \autoref{fig:snba}) as knowledge sources that potentially contain norm behavior. As these sources for norm behavior may require interpretation, knowledge sources to support such an interpretation are needed (s. \emph{knowledge about semantic formalization} in \autoref{fig:snba}). For the German Road Traffic Regulation, national jurisprudential literature, such as monographs, papers, or Article-by-Article Commentary should be used in the context of interpretation. Court cases are of special interest as they provide both norm behavior and an indication of how the law (e.g., the StVO) can be interpreted. 

In our example analysis in \autoref{sec:examp} we use domain-specific methods elaborated in \cite{hildebrand_juristischer_2017, puppe_kleine_2019}. The consequences of this limitation are discussed in \autoref{subsec:crit}. Primarily, we intend to illustrate, how legal knowledge can be integrated as part of the Semantic Norm Behavior Analysis. The process of a legal assessment of a \emph{case} -- which in our case is described by a scenario\footnote{In this case, legal norms that are part of the traffic code are applied to automated vehicles, while they are originally meant to regulate human driving behavior. A fundamental discussion of the applicability of such norms in this context is provided in \cite{gstottner_durfen_2021, steege_automatisierte_2022} from a German legal perspective.} -- is typically structured into four steps \cite{hildebrand_juristischer_2017}. These steps include the \emph{premise}, \emph{definition}, \emph{subsumption}, and \emph{result}\footnote{The translation of these steps is done by the authors. German terms are \emph{Obersatz}, \emph{Definition}, \emph{Subsumtion}, and \emph{Ergebnis}.}. In each of these steps, the case is assessed with respect to relevant legal norms. 
This approach is adopted to apply the Semantic Norm Behavior Analysis with the specific focus on legal analyses in a German target operational domain. 
As a result of applying this approach, legal assumptions and interpretations are systematically documented.

In the \emph{premise}, the key question of the case is formulated \cite{hildebrand_juristischer_2017}. This question shall be answered over the course of the assessment. Good practice \cite{hildebrand_juristischer_2017} demands that the premise stays neutral and efficient, thus avoiding value judgments. The premise shall also provide information that is relevant for answering the case question. Hence, in the formulation of the question, the most important sections are often included.

The second step in a legal assessment is to phrase the \emph{definition} \cite{hildebrand_juristischer_2017}. In the definition, all relevant legal norms and their individual elements of offense are elicited and analyzed to answer the question raised in the premise. 
Based on the considered legal norms, the applicable legal offenses that have implications for the legal consequence in the case are formulated. For all legal offenses, the facts (i.e., features) that need to be true in a case are formulated accordingly. Therefore, if the respective facts are true in a given case, it can be concluded that a legal offense applies to the case. In the context of a scenario-based approach, we use the concept of \emph{facts} to capture semantic concepts in a scene of a functional scenario \cite{menzel_scenarios_2018}. For a specification of target behavior in a logical scenario \cite{menzel_scenarios_2018} it is necessary to define parameter spaces for the identified facts.
While a parameterized specification of facts may be possible, we confine ourselves to the specification of facts (and target behavior) in functional scenarios in the context of this work. As an initial behavior specification must be conducted before system design, we do not apply the Semantic Norm Behavior Analysis to logical and concrete scenarios in this work. A fact in a functional scenario could be, for example, that a \textit{valid pedestrian crossing is captured} or that a \textit{pedestrian intends to use a pedestrian crossing}. Based on the formalized rules, the target behavior can be inferred (e.g., \emph{stopping at a pedestrian crossing}). 

Individual elements of the facts and legal offenses that are part of the definition might be ambiguous (e.g. the concept of a \emph{crossing intention}). 
This particularly applies to undefined legal terms.
Thus, further interpretation of the terms may be necessary to draw conclusions for compliant target behavior in a scenario. 
While there are established methods in the legal domain to support sound interpretation, these methods reach their limits in the case of undefined legal terms \cite{steege_automatisierte_2022}.
Therefore, it is not always possible to achieve a result of an interpretation that is as precise as known from the natural sciences.
On the one hand, such undefined terms guarantee the flexibility of our legal system and enable its further development. On the other hand, they lead to legal uncertainty. Which methods of interpretation are permissible depends in part on the respective area of law. 

In the third step of a legal assessment \cite{hildebrand_juristischer_2017} the \emph{subsumption} is formulated. 
The subsumption includes the specification of target behavior based on the analyzed legal norms.
The core of a subsumption are \emph{syllogistic inferences}. Such inferences allow for the following legal reasoning: If a legal offense is defined to have a legal consequence and if this legal offense applies in a case, then the legal consequence applies in this case. Therefore, syllogistic inferences provide the rationale for why a specified target behavior complies with the considered laws.

Finally, the \emph{result} of the assessment summarizes the answer to the case question that is raised in the premise.

These four steps provide the basis for a traceable documentation of assumptions in the specification of target behavior based on the interpretation of norm behavior. Since this part of the analysis is conducted in natural language, further formalization into a machine-readable format is necessary. In the following section, we show how ontologies can be used to formally capture a behavior specification. This formalization is based on the methods for legal analyses we presented in this section. From an engineering perspective, a formalized approach for behavior specification facilitates traceability of specified behavior to the considered stakeholder needs. However, from a legal perspective, such a formalization is not straightforward. Although formal scientific influences can be found again and again over the course of time, the formalization of the law is not generally deemed permissible in jurisprudence.\footnote{For example, Raimundus Lullus (1235-1315 AD) already had formal scientific influences, such as his approach to combinatorial thinking \cite{steege_automatisierte_2022}.} Thus, we do not claim that legal compliance is guaranteed by applying formalization in the Semantic Norm Behavior Analysis. Instead, the approach enables the traceability of the behavior specification to assumptions, which can be challenged in case the system behaves in a non-compliant manner.

\subsection{Creation of a Target Behavior Model with Ontologies}
 \label{subsec:ontol}
Westhofen~\emph{et~al.} define  \emph{congruence} as a circumstance where the \enquote{eventual goal of the [Automated Driving] System’s development is the implementation of some valid interpretation of the relevant legal concepts}~\cite[p.~14]{westhofen_towards_2022}. However, the authors note that the fundamental idea can be applied to consensus-building in general.
Thus, we adopt their concept of congruence to the consolidation and specification of behavioral requirements based on potentially conflicting stakeholder needs.
Westhofen~\emph{et~al.}~\cite{westhofen_towards_2022} also elaborate key challenges in addressing the congruence problem. We infer that these challenges are caused by inconsistent representations between different domains. 
Therefore, we propose a minimal set of concepts\footnote{In the following, we predominantly use the abstract term \emph{concepts} to pool multiple terms. We only refer to the more detailed terms \emph{entities} and \emph{relations} if necessary. Additionally, an ontology may contain \emph{classes} and \emph{instances} of the entities and relations.} in two ontologies that we identified to be suitable in a behavior specification. We separate the concepts into two ontologies in order to maintain a clear scope of each knowledge representation.

\Figure[ht](topskip=0pt, botskip=0pt, midskip=0pt)[width=0.9\columnwidth]{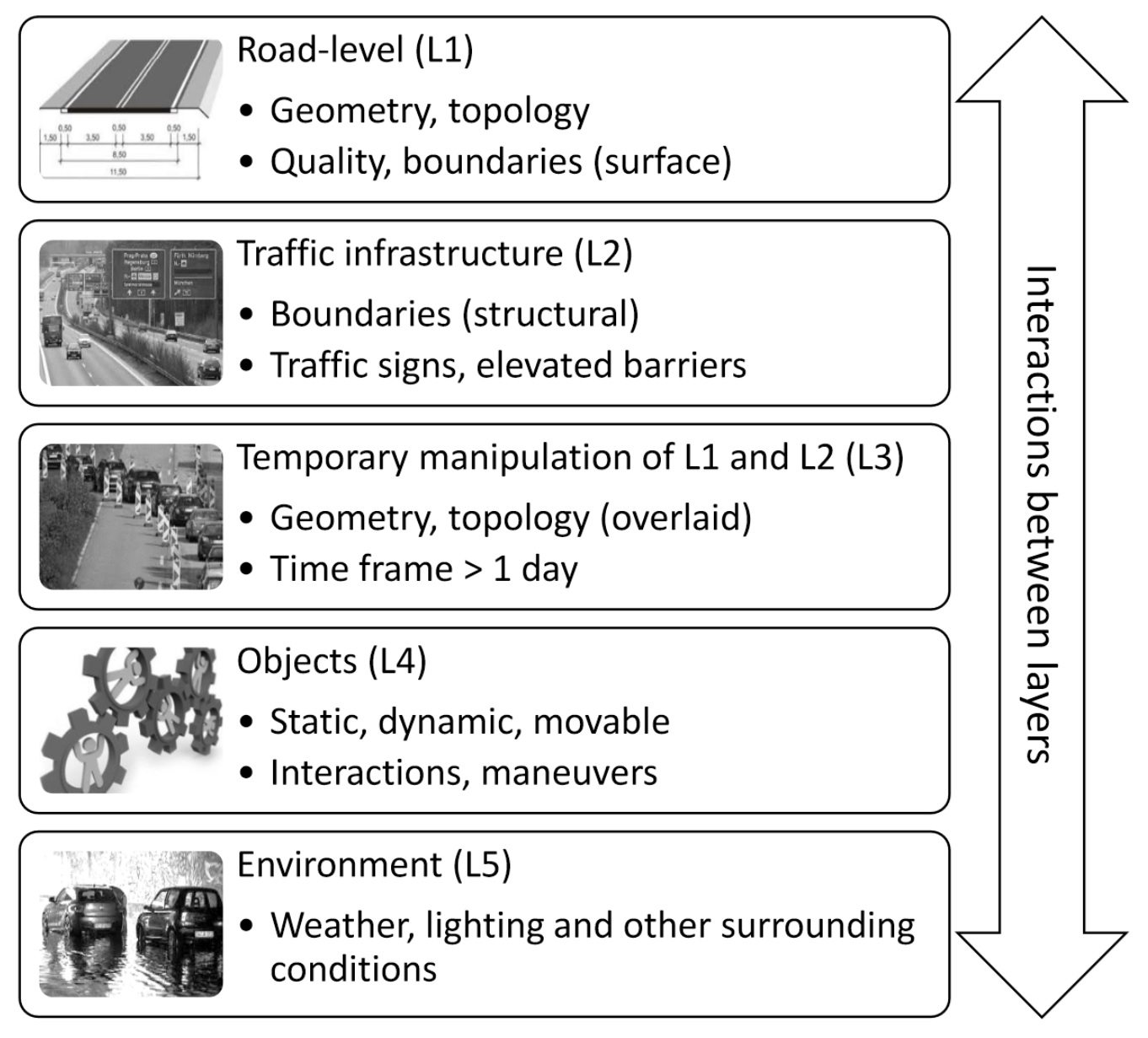}
{Illustration of the 5-Layel-Model. \cite{bagschik_ontology_2018} \label{fig:ebenen}}

First, the operational domain ontology captures the operating conditions in the target operational domain. 
Second, the behavior specification ontology captures behavioral requirements under these operating conditions. 
The operational domain ontology describes the operational domain according to the 5-Layer-Model by Bagschik~\emph{et~al.}~\cite{bagschik_ontology_2018} (neglecting proposed extensions from Scholtes~\emph{et~al.}~\cite{scholtes_6-layer_2021}). The 5-Layer-Model is illustrated in \autoref{fig:ebenen}.
In contrast, the behavior specification ontology establishes traceability between specified behavior and the considered stakeholder needs (which are represented by legal references in this article).
Both ontologies are modeled in the \emph{Web Ontology Language} (OWL), which is based on description logic. The logic programming rules that supplement the ontologies are written in the \emph{Semantic Web Rule Language} (SWRL).

\paragraph{Operational Domain Ontology}
For a scenario-based behavior specification, the scenario-specific context has to be modeled. Such a model should be based on a consistent representation of the operational domain. In our application, entities of the operational domain are describes with the classes \emph{Chronological Entity}, \emph{Scene Entity}, \emph{Parameter}, and \emph{Characteristic} (cf. \autoref{fig:domainonto}).

\emph{Chronological Entity} summarizes classes for the description of the temporal course, like scenario and scene.

The scene elements (\emph{Scene Entities}) contain concepts that are potentially present within a scene and contribute to the understanding of the scene or scenario.
\emph{Scene Entities} include, for example, \emph{L1 Road Level Entities} describing road geometry and topology, such as a \emph{Lane}, or \emph{L4 Movable Object and Behavior Entities} describing road users and their interactions, such as the maneuver \emph{Lane Change}.
However, each \emph{Scene Entity} contains only its essential features.
Thereby, an essential feature is a feature that is shared by all individuals of the respective class.
Features that vary across different individuals of a given \emph{Scene Entity} are called \emph{Characteristics} and are modeled independently.
For example, \emph{Characteristics} include descriptions of different geometric properties, such as \emph{Lane Width Geometry}.
\emph{Lane Width Geometry}, in turn, subsumes different (sub)characteristics, such as \emph{Linear Lane Width} or \emph{Constant Lane Width}.
Each \emph{Scene Entity} is thus qualitatively described by its \emph{Characteristics}.

While we focus on the behavior specification in functional scenarios in this article, the operational domain ontology already accounts for parametric descriptions (i.e. logical and concrete scenarios). Thus, a distinction is made between a \emph{Characteristic} and the \emph{Parameters} describing the \emph{Characteristic}.
While a \emph{Characteristic} describes the features of a scene element qualitatively, a \emph{Parameter} specifies a \emph{Characteristic} quantitatively.
For example, the \emph{Characteristic} \emph{linear lane width} is specified in detail with the \emph{Parameters} \emph{Start Width} and \emph{End Width}.
This explicit separation between \emph{Scenario Entities}, \emph{Characteristics} and \emph{Parameters} supports a scenario description on different levels of abstraction.

\Figure[h!](topskip=0pt, botskip=0pt, midskip=0pt)[width=0.9\columnwidth]{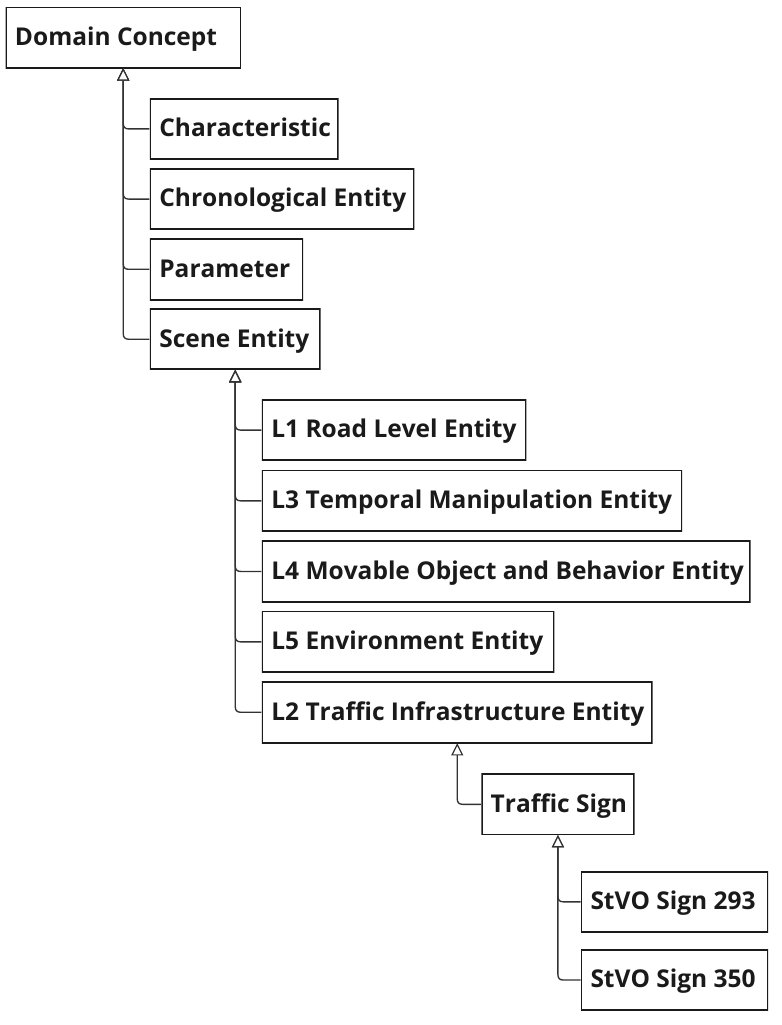}
{Taxonomy of the key concepts in the operational domain ontology.\label{fig:domainonto}}

\paragraph{Behavior Specification Ontology}
The concepts that constitute the behavior specification ontology include 
\emph{Maneuver Option}, 
\emph{Zone},
\emph{Knowledge Source}, 
\emph{Fact},
\emph{Mission}, and 
\emph{Characteristic} (cf. \autoref{fig:behavonto}).

\Figure[h!](topskip=0pt, botskip=0pt, midskip=0pt)[width=0.9\columnwidth]{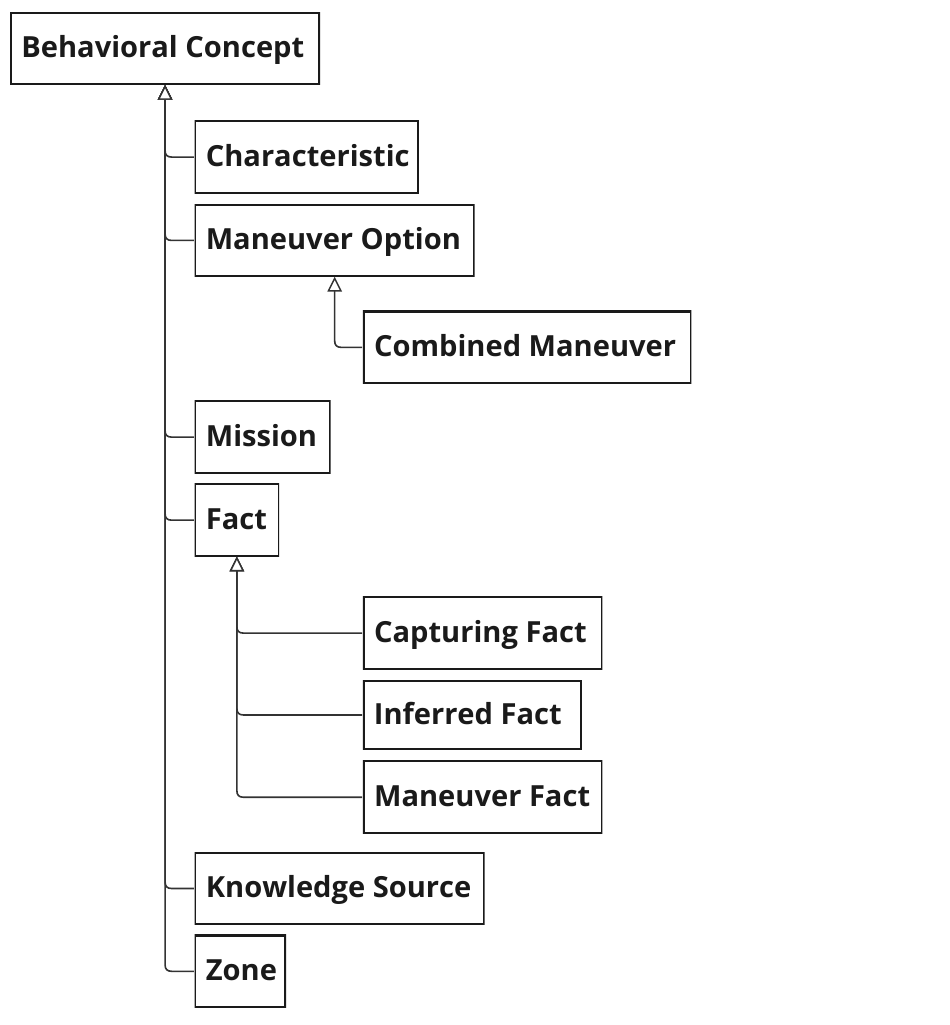}
{Taxonomy of the key concepts in the behavior specification ontology.\label{fig:behavonto}}

As the basic goal of a behavior specification is to specify compliant maneuvers in a scenario, we define abstract \emph{Maneuver Options}. These are combinations of lateral and longitudinal maneuvers. In this example, we adopt maneuvers as proposed by Jatzkowski~\emph{et~al.}~\cite{jatzkowski_zum_2021}. They define \emph{start}, \emph{stop}, \emph{follow desired speed}, and \emph{follow target vehicle} as longitudinal maneuvers; \emph{keep lane}, \emph{change lane}, and \emph{pass} are defined as lateral maneuvers.

\emph{Zones} provide the reference concept for this example of a behavior specification. A zone grid is defined, which serves as a local coordinate system for the ego vehicle. Within this zone grid, scene elements are specified and facts and rules are applied to the modeled scene. As a result, maneuver options are inferred based on the scene elements in the zone grid and the applied facts and rules. This approach is comparable to the zone graph introduced by Butz~\emph{et~al.}~\cite{butz_soca_2020} and Bagschik~\cite{bagschik_systematischer_2022}. The authors define zones for regions of the scenery. As a result, multiple similar sceneries can be described by one zone graph. In contrast to Butz~\emph{et~al.}~\cite{butz_soca_2020} and Bagschik~\cite{bagschik_systematischer_2022} we do not focus on the abstraction of multiple sceneries. Instead, we define one persistent zone grid to which the facts and rules are applied. This results in an iterative refinement of the zone grid until it can be applied to a set of suitable scenarios. As this example is limited to two scenarios, the zone grid presented in \autoref{fig:zone} is relatively minimal.

\Figure[h!](topskip=0pt, botskip=0pt, midskip=0pt)[width=0.9\columnwidth]{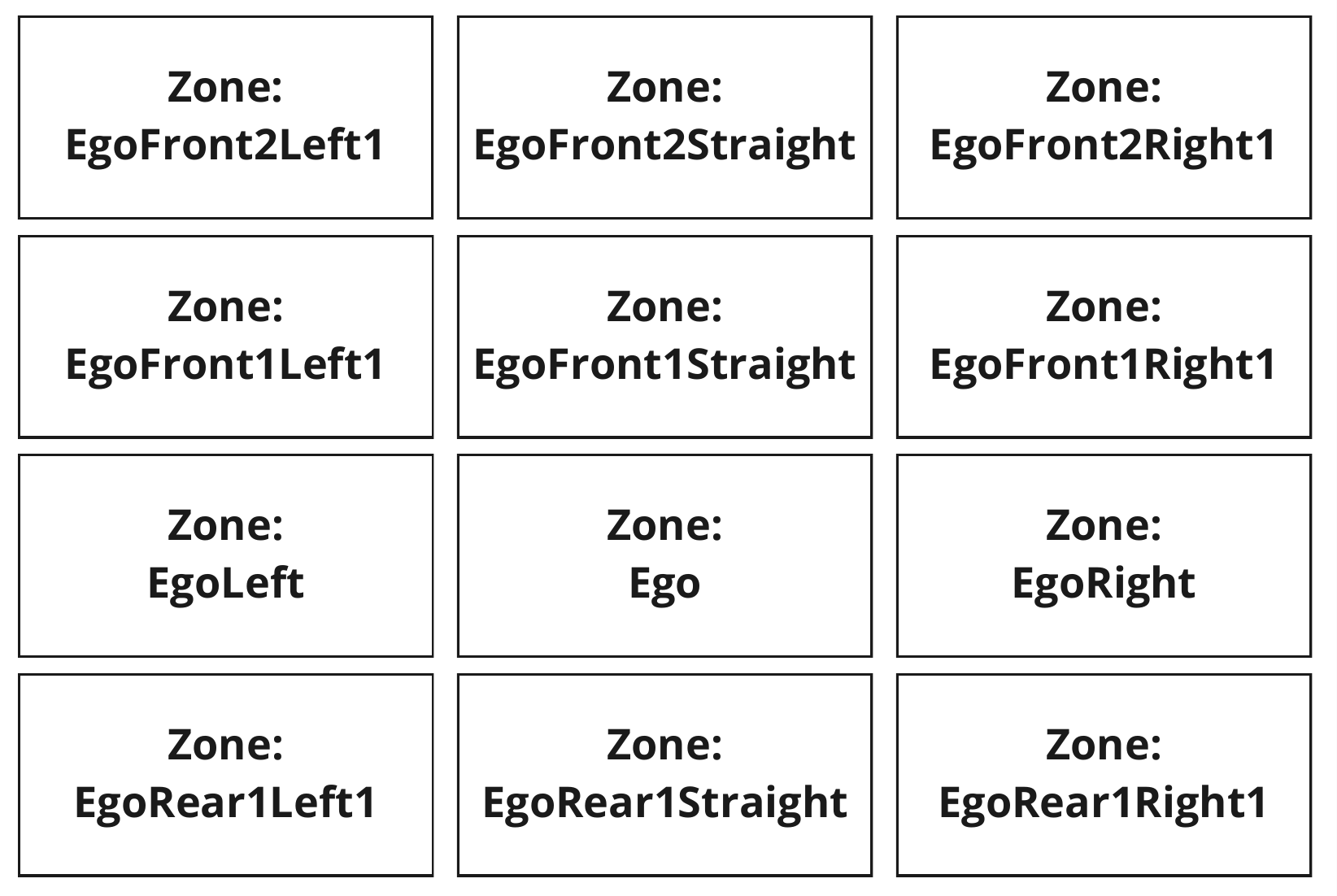}
{Zones are used as an interface between the model of the operational context and the behavior specification. Entities of the scenario-specific context are defined to be in a zone. Subsequently, the behavior specification refers to the entities as part of the zone grid. \label{fig:zone}}

\emph{Knowledge Sources} are the root of the behavior specification as they capture the stakeholder needs towards the behavior of an automated vehicle. In this article, we focus on the German Road Traffic Regulation as a knowledge source. In \autoref{subsec:domainan} we described domain-specific methods (i.e. methods to analyze legal texts in Germany). As a result of applying such methods, it is possible to model the rationale for the specification of a maneuver option in the ontology by referring from a maneuver to the respective knowledge source.

In the Semantic Norm Behavior Analysis, we propose to analyze knowledge sources with respect to relevant facts that determine the specified behavior. The \emph{Fact} class in the ontology should be specified based on the concrete facts that were defined in the previous analysis step (s. \autoref{subsec:domainan}). Facts are the foundational concept that is taken from the Phenomenon-Signal Model \cite{beck_phenomenon-signal_2022}. As facts are used in a phenomenological sense, facts shall be related to scene elements. Therefore, facts capture the causal relations between the scenario-specific context and the specified maneuver options\footnote{The distinction between capturing facts, inferred facts, and maneuver facts is added to provide further structure to the behavior specification but is not necessary.}.

To specify the behavior of an automated vehicle in a scenario, a respective agent (i.e., an ego-vehicle) needs to be selected. The ego-vehicle and its context are specified in a starting scene. To specify behavior that leads to a desired end scene a \emph{Mission} needs to be defined. The mission specifies the goal that is pursued by the ego-vehicle. It determines which maneuver options can be taken based on the specified set of facts and rules.

\paragraph{Behavioral Rules}
The previously specified ontologies are represented in the Web Ontology Language (OWL), which is based on first-order description logic. Additionally, we use logic programming rules to complement reasoning capabilities in the ontologies. These rules are formalized in the Semantic Web Rule Language (SWRL). 
Both languages are subject to the open world assumption, which causes uncertainty for all statements that are not included in the specification. If a fact or a rule is not modeled, it does not mean that it is definitely false. This property of the modeling language is in line with the general challenge of developers, who have to define an explicit specification for the conditions they considered: Conditions that were not considered by the developers, but that are relevant for vehicle behavior constitute an incomplete behavior specification. In this respect, the open world assumption in OWL and SWRL do not constitute a direct limitation: Modeled knowledge can be checked for consistency, the \enquote{completeness} of the knowledge must be validated on a process level. For example, safety analyses can assist in uncovering and treating cases, where the reasoning result is not consistent with the expected outcome of the specification (e.\,g. no possible maneuver is inferred). General limitations with respect to verification and validation of the behavior specification are discussed in the following section (\ref{subsec:vv}).

In the context of a behavior specification, we propose to use SWRL-rules as behavioral rules. As a result, causal relations between scene entities, facts and maneuver options can be inferred by the reasoner. Behavioral rules follow the simplified structure specified in \autoref{lst:swrl_gen}.

\begin{figure}
    \centering
    \begin{lstlisting}[style=swrl, caption={Abstract rules written in the Semantic Web Rule Language (SWRL) to infer applicable facts and maneuver options.}, label={lst:swrl_gen}]
Scene_Entity(?e) ^ Zone(?z) ^
is_in(?e,?z) ^ Perception_fact(?pf)
-> fact_applies(?pf,true)

Fact(?f1) ^ Fact(?f2) ^ 
fact_applies(?f1,true) ^ 
fact_applies(?f2,true) ^ Fact(?f3)
-> fact_applies(?f3,true)

Fact(?f) ^ fact_applies(?f,true) ^
ManeuverOption(?m)
-> maneuver_applies(?m,true)
\end{lstlisting}
\end{figure}

Behavioral rules are formalized according to the results of the first step in the Semantic Norm Behavior Analysis (s. \autoref{subsec:domainan}). Facts that apply to a scene are inferred by the reasoner, if a scene contains scene entities (in the operational domain ontology) and they are modeled in a relevant zone of the zone grid (in the behavior specification ontology). Based on the inferred facts, the behavioral rules specify maneuver options that are compliant with the behavior specification. 
This way of deductive reasoning is especially useful at design time, where the \enquote{known scenarios} (according to ISO~21448)~-- and thus the known operating conditions~-- are defined by the developer. As a result, the modeling language does not need to represent uncertainties about the perception system or the behavior of other traffic participants. The behavior specification consists of derived requirements. The need for uncertainty representation at runtime can be an outcome of these requirements. Such uncertainties need to be addressed in the technical implementation of the automated vehicle as it has to make decisions under uncertain conditions. However, developers first have to make assumptions and specify requirements under determined conditions.

\subsection{Verification and Validation of the Behavior Specification}
\label{subsec:vv}
After an analysis of knowledge sources and a specification of target behavior, the behavior specification needs to be verified and validated. In the Semantic Norm Behavior Analysis, target behavior is modeled using description logic and logic programming. Thus, in the scope of the Semantic Norm Behavior Analysis, verification of a behavior specification can be supported by proving formal consistency of the modeled concepts with a reasoner to show that the specification is unambiguous. However, this does not necessarily mean that the modeled target behavior is \emph{correct}. To prove correctness of target behavior, the specification also needs to be \emph{complete}. However, this poses multiple challenges. 

First, target behavior is specified for a set of scenarios. 
In the scope of these scenarios, a behavior specification is axiomatic. That is, behavior can be specified as a consistent, closed set of requirements. However, the operational environment of an automated vehicle is an open world. This leads to inherent uncertainties in the description of the operational environment - in our case modeled in the operational domain ontology. 
As behavior is specified with respect to entities in the operational environment, the validity of the specified axioms is limited by the completeness of the operational domain ontology. Therefore, a behavior specification can only be formally verified regarding the consistency of the specified axioms. The axioms themselves and the description of the operational environment thus still require validation. From a safety perspective, the inherent uncertainty about residual specification insufficiencies leads to an inherent risk that can be reduced but not eliminated \cite{nolte_representing_2018}. 

Additionally, validation of the scenario-based behavior specification shares validation challenges with any scenario-based approach. While scenarios can be used to structure the operational context and behavior of an automated vehicle, a proof of a sufficiently valid scenario catalog is yet to be given. 
As the Semantic Norm Behavior Analysis follows a scenario-based approach, the axioms composing the behavior specification require validation with respect to the operational environment.
One possible measure to address the incompleteness of the behavior specification after an automated vehicle is deployed is an ongoing monitoring of the automated vehicle throughout the product lifecycle to identify specification insufficiencies. For example, ISO~21448~\cite{noauthor_road_2022-1} provides guidelines for operation phase activities to ensure the safety of the intended functionality during operation.
We argue that systematic field monitoring is supported by the explicit specification of behavior and the operational environment. As assumptions about target behavior and the operational environment are explicitly documented, data that refutes these assumptions can be used to refine the behavior specification. 

Finally, validation of the specified target behavior regarding the considered knowledge sources poses additional challenges. 

First, the selected knowledge sources need to sufficiently cover behavior-related stakeholder needs in the target operational domain. As a result, knowledge sources require validation by domain experts. For example, when legal sources such as the German Road Traffic Regulation are defined as a source for the behavior specification, legal experts need to validate whether this regulation can be taken as an exclusive source for behavior specification.

Second, compliance of the specified target behavior with the selected knowledge sources needs to be validated. For individual scenarios, an assessment is necessary whether stakeholder needs are appropriately consolidated by the speci\-fied behavioral requirements -- especially when conflicts occur in the considered needs. In the example of a legal assessment of target behavior, case law provides guidelines for previous interpretations of the German Road Traffic Regulation.

\section{Example Behavior Specification}
\label{sec:examp}
In this section we provide an example of how the Semantic Norm Behavior Analysis can support a traceable specification of target behavior in two scenarios based on an excerpt of the German Road Traffic Regulation. These scenarios show that even with a limited number of scene elements, a behavior specification requires interpretations that lead to residual uncertainties.

Figures \ref{fig:examp1} and \ref{fig:examp2} show two functional scenarios where the automated ego-vehicle is entering a T-crossing from the left. In scenario \emph{A} (\autoref{fig:examp1}) a pedestrian is attempting to use the pedestrian crossing at the T-crossing by walking directly onto it. In scenario \emph{B} (\autoref{fig:examp2}) a pedestrian does not use the marked area of the pedestrian crossing, but is crossing the road between the ego-vehicle and the pedestrian crossing.

\Figure[h!](topskip=0pt, botskip=0pt, midskip=0pt)[width=0.9\columnwidth]{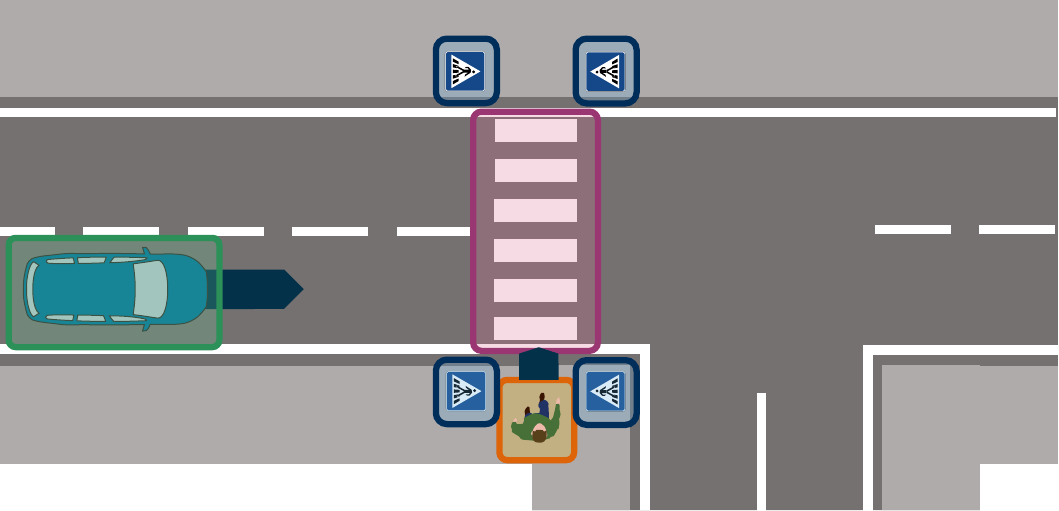}
{Scenario A: A pedestrian attempts to use a pedestrian crossing. The blue ego-vehicle approaches the pedestrian crossing from the left. Sign 350 (blue signs) and sign 293 (marking on the road) signalize the pedestrian crossing.\label{fig:examp1}}

\Figure[h!](topskip=0pt, botskip=0pt, midskip=0pt)[width=0.9\columnwidth]{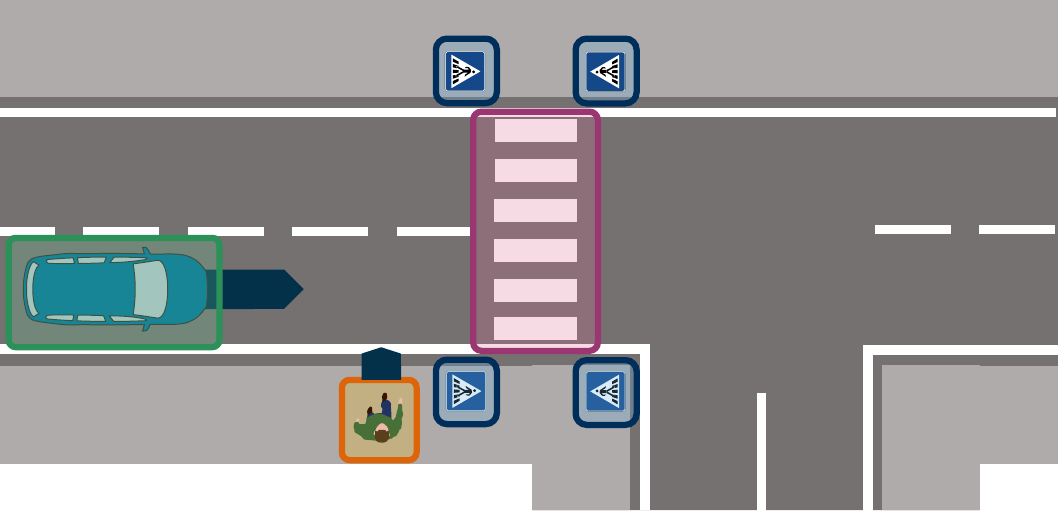}
{Scenario B: In contrast to Scenario A, the pedestrian attempts to cross the road between the pedestrian crossing and the ego-vehicle. As for Scenario A the blue ego-vehicle approaches the pedestrian crossing from the left. Sign 350 (blue signs) and sign 293 (marking on the road) signalize the pedestrian crossing.\label{fig:examp2}}

As we follow the procedure described in the Semantic Norm Behavior Analysis, we first analyze norm behavior based on selected knowledge sources of stakeholder needs (\autoref{subsec:legal}). 
We then show a formalization of target behavior using ontologies in \autoref{subsec:formal}. Furthermore, we propose representations of the behavior specification to support communication with different stakeholders in 
\mbox{\autoref{subsec:repres}}. In \autoref{subsec:eval} we evaluate the results of this example application of the Semantic Norm Behavior Analysis.

\subsection{Application of Legal Methods in the Semantic Norm Behavior Analysis}
\label{subsec:legal}
In this example, the knowledge sources \S 26 of the German Road Traffic Regulation (StVO) and the according administrative guideline\footnote{German: \emph{Verwaltungsvorschrift}} for \S 26 StVO provide the basis of the analysis as they specifically address pedestrian crossings (s. \emph{Step~1} in \autoref{fig:snba}). Consequences of this limitation are discussed in \autoref{subsec:crit}. We give this example to show an application of the Semantic Norm Behavior Analysis.

According to the method description in \autoref{subsec:domainan} the first step of this analysis is the specification of a premise. Usually, the premise is formulated as a closed question. Since the goal of the assessment in the case of the Semantic Norm Behavior Analysis is to identify the key facts and rules that determine compliant behavior, we deliberately phrase the premise as an open question.

Given the considered knowledge sources and the scenarios a premise in this case could be:\\

\noindent
\textbf{Premise:}
\begin{itemize}
    \item Which duties apply to the driver of the ego-vehicle according to \S 26 StVO in the scenarios depicted in Figures \ref{fig:examp1} and \ref{fig:examp2}?
\end{itemize}

\vspace{3mm}

In the definition, we state that \S 26 StVO refers to the legal offense that a \emph{pedestrian crossing is present}. If this legal offense applies in a case, \S 26 StVO specifies that the driver of a vehicle has the duty to yield right of way to pedestrians. To assess the applicability of the legal offense, the operational context needs to be analyzed based on the administrative guideline of the StVO. The guideline defines that a \emph{sign 293} is obligatory for a valid marking of a pedestrian crossing. Additionally, \emph{sign 350} indicates the presence of a pedestrian crossing. As a result, both the German Road Traffic Regulation and the administrative guideline provide a baseline for the specification of target behavior. However, for a consistent specification of behavioral requirements, facts and rules need to be formalized. For example, \S 26 StVO includes the condition of a \enquote{foreseeable crossing intention} to define the right of way for a pedestrian. Taken out of context, this term is ambiguous and needs to be specified for the considered scenarios in the \emph{subsumption}.

As we conducted this case study, we found that the \emph{definition} especially supports the interdisciplinary communication required in the context of behavior specification. More specifically, the definition of \emph{facts} provides a common ground for legal experts and engineers. From a legal perspective, facts need to be defined in order to assess the applicability of a legal offense in a certain context. From an engineering perspective, facts provide a causal link between entities in the operational environment and specified maneuvers. 

Due to the limited scope of this example analysis, a possible part of the definition is:\\

\noindent
\textbf{Definition:}
\begin{itemize}
    \item According to \S 26 StVO the driver of a vehicle has the duty to yield right of way to pedestrians, which foreseeably intend to use a pedestrian crossing and enable them to cross the road.
    \item According to the administrative guideline of \S 26 StVO a pedestrian crossing is marked by traffic sign 293 (i.e., lines on the road). Additionally, sign 350 indicates the presence of a pedestrian crossing.
\end{itemize}

\vspace{3mm}

As a result, the relevant sources, facts, and terms are compiled in the definition. Subsequently, we apply the definition to the specified scenarios and ask for the applicable legal consequences as part of the subsumption. One example of a syllogistic inference is that the existing signs are meant to signalize a pedestrian crossing. 
\\

\noindent
\textbf{Subsumption:}
\begin{itemize}
    \item The scenarios described in Figure \ref{fig:examp1} and \ref{fig:examp2} contain sign 293. Additionally, sign 350 indicates a pedestrian crossing. According to the administrative guideline for \S 26 StVO both signs signalize the presence of a pedestrian crossing.
    \item A pedestrian is present in both scenarios. The position of the pedestrian is defined to be near the pedestrian crossing. We assume that this fact signalizes the pedestrian's crossing intention.
    \item The ego-vehicle is near the pedestrian crossing.
    \item As a result, the ego-vehicle must stop at the pedestrian crossing to yield to the pedestrian.
\end{itemize}

\vspace{3mm}

Finally, the result of the assessment summarizes the answer to the case question that is raised in the premise, given a set of assumptions. These assumptions include that the scenario is defined in a closed community (i.e., a city or village) as this is a prerequisite for a valid pedestrian crossing in Germany. Further assumptions are the noticeable crossing intention of the pedestrian in both scenarios, which we assume to be based on a position close to the pedestrian crossing. While this assumption may be sensible in the considered scenarios, such an assumption would likely have to be revisited in a larger scale setting.
Therefore, to obtain legally valid results, a more thorough analysis of the German Road traffic Regulation, further legal sources, such as publications and court decisions would have to be conducted as part of the interpretation of law.
\\

\noindent
\textbf{Result:}
\begin{itemize}
    \item The driver of the ego-vehicle in Figure \ref{fig:examp1} and \ref{fig:examp2}, according to \S 26 StVO, has the duty to enable the pedestrian to cross the road at the pedestrian crossing. Therefore the driver must drive with moderate speed and stop at the pedestrian crossing if necessary.
\end{itemize}

\vspace{3mm}

This example application of legal methods shall highlight the potential for interdisciplinary work within a behavior specification by following the Semantic Norm Behavior Analysis. We do not claim that the selected methods are applicable in other legal contexts. The consequences of this limitation are discussed in \autoref{subsec:crit}. 

\subsection{Formalization of Facts and Rules in an Ontology}
\label{subsec:formal}

The example of a legal analysis provides a basis for the formalization of target behavior, as it explicitly documents the rationale for the behavior specification. In this section, we use an ontology (s. \emph{Step~2a} and \emph{2b} in \autoref{fig:snba}) to formally represent this rationale. 

\begin{table*}[htbp]
    \caption{Description of classes in the operational domain ontology and the behavior specification ontology that are relevant for the considered example.}
    \label{tab:entities}
    \centering
    \setlength{\tabcolsep}{12pt} 
    \begin{tabular}{p{6.5cm}p{6.5cm}} 
        \toprule 
        \textbf{Class name in Operational Domain Ontology} & \textbf{Description}\\
        \midrule 
        \texttt{Sign293} & Lines on the road marking the area of a pedestrian crossing\\
        \texttt{Sign350} & Blue sign indicating a pedestrian crossing\\
        \texttt{TrafficParticipant:CarWithAgent} &  The ego-vehicle driving towards the pedestrian crossing with the desired speed\\
        \texttt{TrafficParticipant:Pedestrian} & A pedestrian standing at the pedestrian crossing\\
        \midrule 
        \textbf{Class name in Behavior Specification Ontology} & \textbf{Description}\\
        \midrule 
        \texttt{Zone:EgoFront2Straight} & Zone where sign 293 is allocated\\
        \texttt{Sign293\_captured} & Fact that sign 293 is captured by the ego-vehicle\\
        \texttt{Sign350\_captured} & Fact that sign 350 is captured by the ego-vehicle\\
        \texttt{ValidPedestrianCrossing} & Fact that a pedestrian crossing is signalized\\
        \texttt{EgoPositionNearPedestrianCrossing} & Fact that the ego-vehicle is near a pedestrian crossing\\
        \texttt{PedestrianNearPedestrianCrossing} & Fact that a pedestrian is near a pedestrian crossing\\
        \texttt{PedestrianCrossingIntention} & Fact that a pedestrian intends to cross the road at a pedestrian crossing\\
        \texttt{StopToEnableCrossing} & Fact that the ego-vehicle shall yield to the pedestrian by stopping at the pedestrian crossing\\
        \texttt{KeepLane\_FollowDesiredSpeed} & Combined maneuver of the lateral maneuver lane keeping and the longitudinal maneuver following desired speed\\
        \texttt{KeepLane\_Stop} & Combined maneuver of the lateral maneuver lane keeping and the longitudinal maneuver stopping\\
        \texttt{FollowRoad} & Mission of the ego-vehicle to follow the road\\
        \bottomrule 
    \end{tabular}
\end{table*}

The behavior specification ontology refers to concepts introduced in \autoref{subsec:ontol} and represents the facts and rules that we analyzed in the previous section.  
Additionally, the operational domain ontology represents the entities that are part of the operational context. A scenario description based on the operational domain ontology is composed of multiple entities that are not relevant to the facts that we present in this example. Hence, we confine ourselves to the key instances that are relevant for this specification of target behavior (s. \emph{Step~2a} in \autoref{fig:snba}). A more detailed description of concepts in the ontologies can be found in \autoref{tab:entities}.

Relevant entities from the traffic infrastructure (i.e. layer two of the 5-Layer-Model) in the operational domain ontology are sign 293 and sign 350. On layer four, movable objects are represented. In this example, these objects are the ego-vehicle and a pedestrian.

In the behavior specification ontology, the most important instances are two maneuver options. Either the ego-vehicle stays in its current maneuver (keeping its lane and following the desired speed) or it stops. 

Based on these entities and their relations, we define SWRL-rules (\autoref{lst:swrl_examp}) that formally capture the results of our example legal analysis (s. \emph{Step~2b} in \autoref{fig:snba}). For example, the rules include statements such as: If there is an instance \verb|e| of sign 293, there is an instance \verb|z| of the zone in front of the ego-vehicle, \verb|e| is positioned in \verb|z|, and there is instance \verb|pf| of the fact that sign 293 shall be captured, then the fact \verb|pf| applies (rule 1). The other example rules follow the same syntax. They lead to the inference that there is a valid pedestrian crossing (rule 2), the pedestrian has a crossing intention (rule 3), and that the ego-vehicle should perform a stopping maneuver (rule 4).

\begin{figure}
    \begin{lstlisting}[style=swrl, caption={Subset of SWRL-rules for facts and maneuver options specified for the example scenarios.}, label={lst:swrl_examp}]
# Rule 1
Sign293(?e) ^ Zone:EgoFront2Straight(?z) ^
is_in(?e,?z) ^ Sign293_captured(?pf)
    -> fact_applies(?pf,true)

# Rule 2
Sign293_captured(?f1) ^ 
Sign350_captured(?f2) ^ 
EgoPositionNearPedestrianCrossing(?f3) ^ 
fact_applies(?f1,true) ^ 
fact_applies(?f2,true) ^ 
fact_applies(?f3,true) ^ 
ValidPedestrianCrossing(?f4)
    -> fact_applies(?f4,true)

# Rule 3
ValidPedestrianCrossing(?f1) ^ 
PedestrianNearPedestrianCrossing(?f2) ^ 
fact_applies(?f1,true) ^ 
fact_applies(?f2,true) ^ 
PedestrianCrossingIntention(?f3)
    -> fact_applies(?f3,true)

# Rule 4
PedestrianCrossingIntention(?f1) ^ 
fact_applies(?f1,true) ^ 
KeepLane_Stop(?m) 
    -> maneuver_applies(?m,true)
\end{lstlisting}
\end{figure}

This inference is in line with our analysis in \autoref{subsec:legal}. In a scenario where there is no pedestrian near the pedestrian crossing, the compliant maneuver option would be to perform the maneuver of following the desired speed. 

In addition to the mere inference of a compliant maneuver, the formalized rules can explicitly be linked to assumptions in the norm behavior analysis. These assumptions can be distinguished by assumptions about the relevant concepts in the scenario (rule 1 and 2) and assumptions about the behavior of other agents in the scenario (rule 3 and 4). These assumptions are made in the analysis of knowledge sources. For example, in our legal analysis of the German Road Traffic Regulation we assumed that the fact of a pedestrian position being close to a pedestrian crossing signalizes their crossing intention. We discuss issues resulting from this assumptions in the following section (\ref{subsec:repres}).

In this minimal example, it already becomes evident that further details in the behavior specification ontology will require a more detailed description of the operational context. The fact of a pedestrian position being near a pedestrian crossing, for example, is specified without parametric constraints and solely depends on the topology of the specified zones. 

Throughout the design and application of the Semantic Norm Behavior Analysis, we faced the challenge of clearly communicating the content of the behavior specification. Therefore, in the following section, we present representations that were particularly helpful in the communication with different domain-experts.

\subsection{Representations for the Communication of an Ontology-based Behavior Specification}
\label{subsec:repres}
In the context of this case study example, we mainly needed to communicate with legal experts and system engineers. First, legal experts needed to be consulted for the conducted legal analysis and were indispensable for the elicitation of facts and rules. Second, systems engineers needed to understand the specified behavior to derive technical requirements and to provide feedback on the feasibility of specified maneuvers.

For the communication with legal experts, a semi-formal representation has proven valuable: \emph{Causal Behavior Graphs} are directed acyclic graphs that model causal relations between context elements and specified maneuvers. They are an abstract, semi-formal representation of the facts and rules that constitute the behavior specification in a considered scenario.

\Figure[h!](topskip=0pt, botskip=0pt, midskip=0pt)[width=1.7\columnwidth]{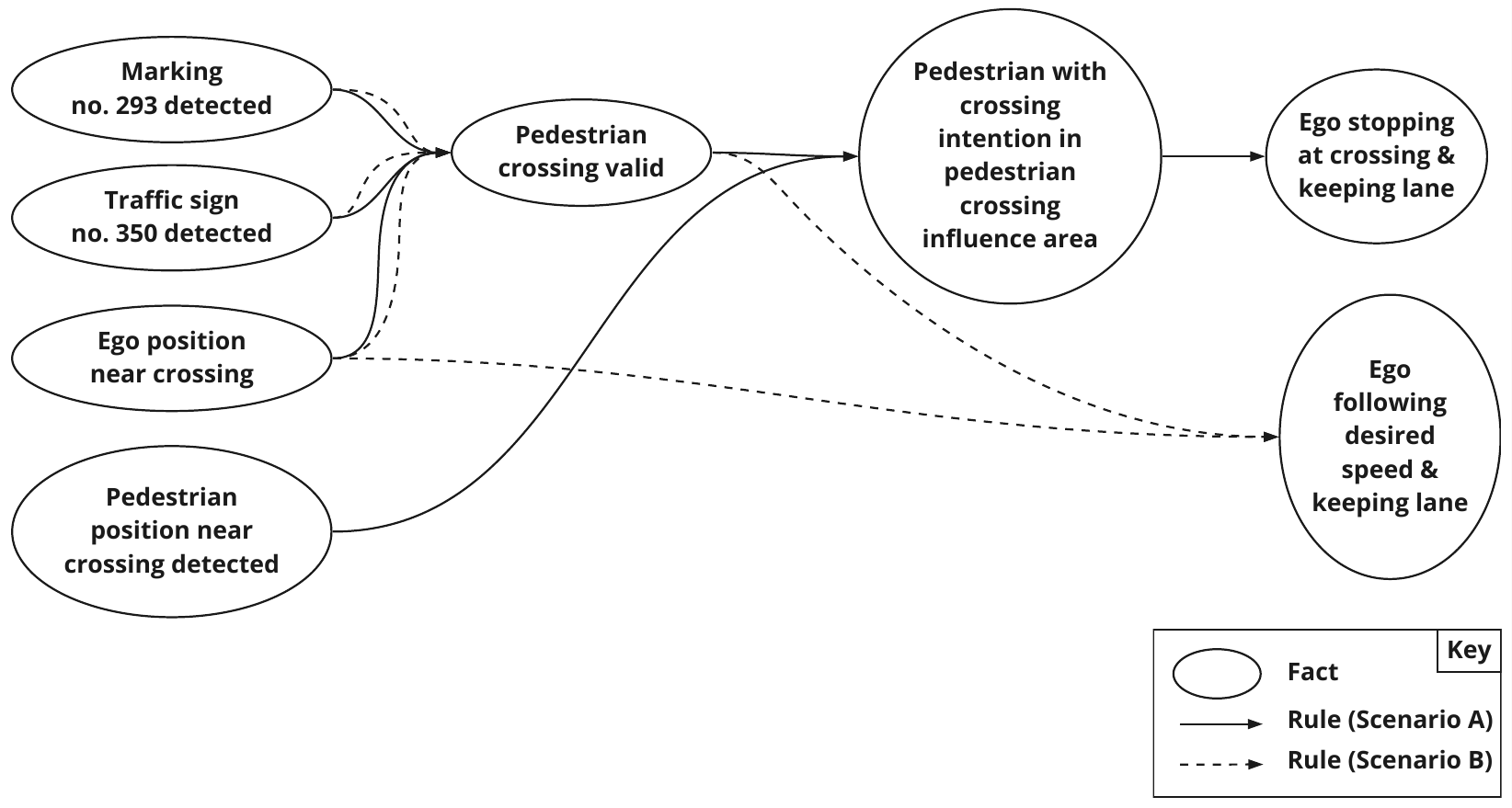}
{The \emph{Causal Behavior Graph} represents the facts and rules from the ontology-based behavior specification in a simplified manner to facilitate interdisciplinary communication. Nodes represent facts whereas edges represent causal relations as modeled by the respective rules.\label{fig:graph}}

\autoref{fig:graph} depicts the Causal Behavior Graph (CSG) for both considered scenarios. Nodes of a Causal Behavior Graph describe instances of facts in the behavior specification. Edges display the causal relation between the facts as formalized in the rules of the ontology. 
Our experience in discussing with experts from legal sciences  is that Causal Behavior Graphs provide a useful tool in a bidirectional communication. On the one hand, Causal Behavior Graphs enable ontology engineers to present the behavior specification to legal experts. On the other hand, the graph-based representation supported the elicitation of facts and rules in order to include them in the ontology.

Additionally, we needed to communicate the contents of the ontology-based behavior specification to systems engineers. 
However, languages designed for ontology engineering, for example, OWL and SWRL do not specify guidelines for visual representations. 
In the case study presented in this article, we found that visualizations support the communication of a behavior specification.
To model visual representations of a behavior specification, alternative languages are needed that at the same time do not sacrifice traceability of the formal concepts. Jenkins~\emph{et~al.}~\cite{jenkins_semantically-rigorous_2012} elaborate that a combination of OWL and SysML regarding modeling complex systems can leverage the formal reasing of OWL and graphical notation of SysML.
Meyer~\emph{et~al.}~\cite{meyer_scenario-_2022} also propose to use SysML in a model-based systems engineering procedure. As Meyer~\emph{et~al.}~\cite{meyer_scenario-_2022} do not explicitly model causal relations between context elements and driving maneuvers, we extend their application of SysML with an alternative metamodel. This metamodel follows the concepts of the Phenomenon-Signal Model \cite{beck_phenomenon-signal_2022} and uses \emph{facts} as a key concept to refer from the specified behavior to context elements in the considered scenario.

\Figure[h!](topskip=0pt, botskip=0pt, midskip=0pt)[width=1.9\columnwidth]{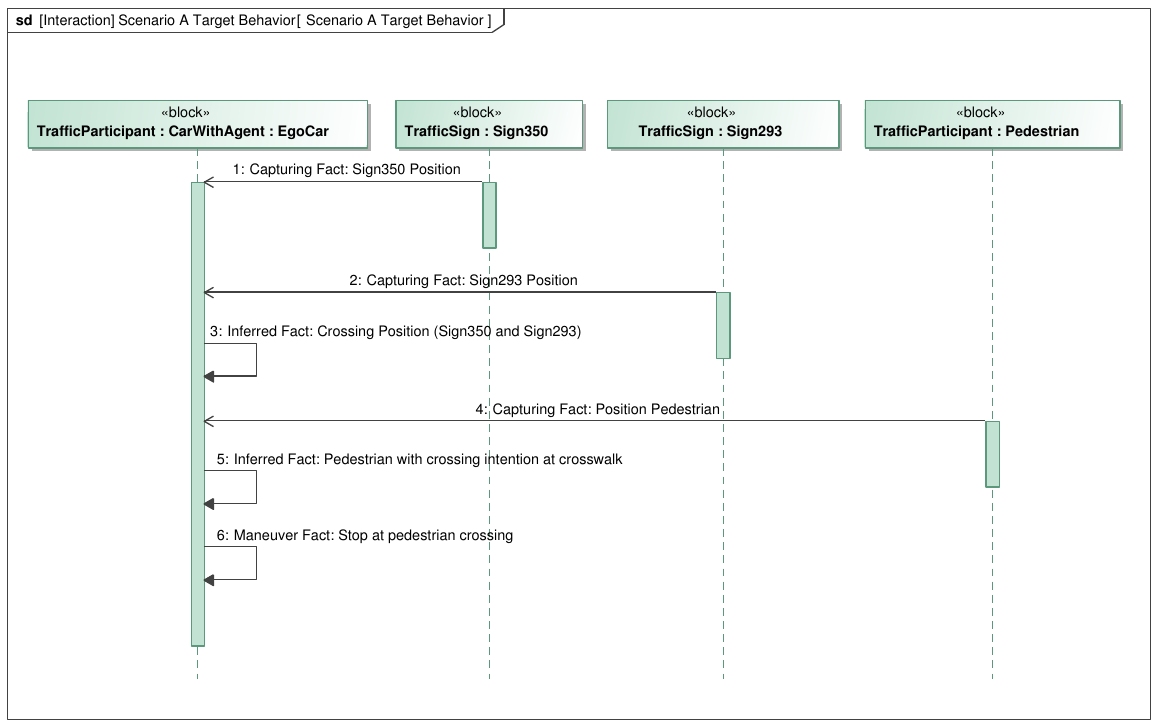}
{A SysML sequence diagram of the example ontology-based behavior specification modeled in \textit{CATIA Magic Cyber-Systems Engineer 2024}. Object lifelines are instances in the operational domain ontology. Messages are facts modeled in the behavior specification ontology. A fact that is directly causally related to an object, which is not the ego-vehicle is modeled as a message between the respective object and the ego. Facts that are inferred based on other facts are self-referenced messages from the ego-vehicle.\label{fig:seq}}

In this example, we show how a behavior specification in OWL can be represented as a sequence diagram in SysML. \autoref{fig:seq} depicts the sequence diagram that is modeled according to our example of an ontology-based behavior specification.
As a result, we obtain a representation of the behavior specification in an established Systems Engineering modeling language instead of OWL. In this representation, we sacrifice visual information regarding the causality between facts. However, a SysML-model can be used to establish formal traceability between architectural elements of an automated vehicle and its behavior specification. Such explicit connections between specified behavior and, for example, functional elements were proposed by Reschka~\cite{reschka_fertigkeiten-_2017} and Nolte~\emph{et~al.}~\cite{nolte_towards_2017}. Meyer~\emph{et~al.}~\cite{meyer_scenario-_2022} propose sequence diagrams to model specified behavior in a scenario. However, the metamodel is not presented by the authors. In contrast, the objects in the sequence diagram shown in \autoref{fig:seq} are directly derived from elements of the operational domain ontology. Messages between objects are modeled based on facts and rules from the behavior specification ontology.

\subsection{Scenario-based Evaluation of the Example Behavior Specification}
\label{subsec:eval}
In this example case study, we showed an application of the Semantic Norm Behavior Analysis. As a result, we obtained a behavior specification that is based on a legal analysis including explicitly documented assumptions. Since the behavior specification is formally described, we can already confirm that the specification is consistent. However, we still need to assess, whether the formal concepts in the ontology-based behavior specification satisfy the expected behavior for the example scenarios (s. \emph{Step~3} in \autoref{fig:snba}).

Throughout this example (\autoref{sec:examp}) we used the same behavior specification for both considered scenarios (Figure \ref{fig:examp1} and \ref{fig:examp2}). If we apply the specified SWRL-rules for the first scenario, the reasoner infers that the only compliant maneuver option is a stopping maneuver by the ego-vehicle in the zone before the pedestrian crossing. In the first scenario, the pedestrian stands in the zone next to the pedestrian crossing. However, in the second scenario, the pedestrian stands on the sidewalk between the pedestrian crossing and the ego-vehicle. Under the conditions of the second scenario, the fact that a pedestrian is near a pedestrian crossing is not inferred to be true. As a result, the maneuver option for the ego-vehicle is inferred to be a lane keeping maneuver combined with following its desired initial speed. In this scenario, this leads to a contact between the pedestrian and the ego-vehicle. 
Therefore, an adaptation of the behavior specification is necessary.

\Figure[h!](topskip=0pt, botskip=0pt, midskip=0pt)[width=1.7\columnwidth]{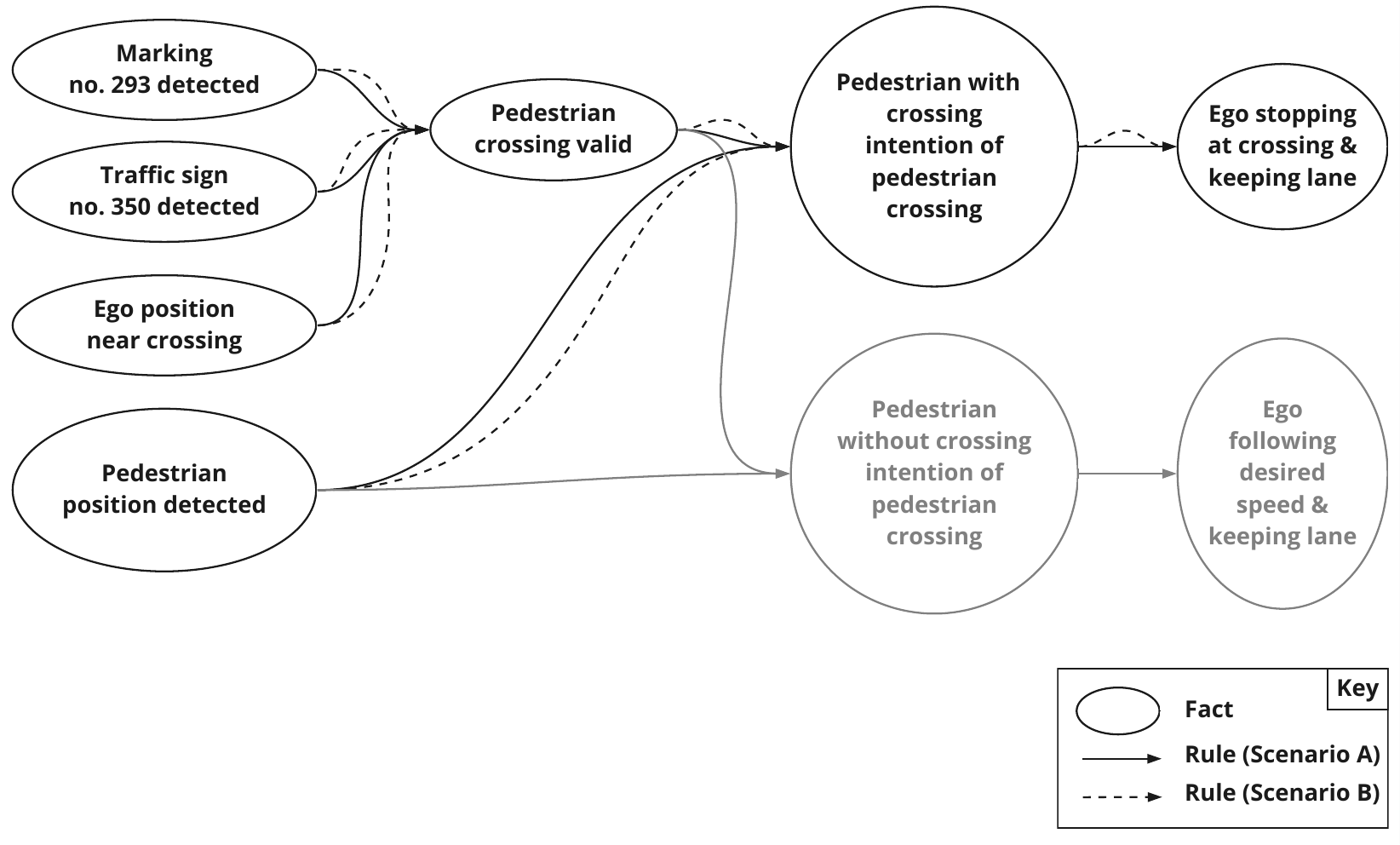}
{The Causal Behavior Graph shows that in both considered scenarios the inferred maneuver is a stopping maneuver. This is due to the treatment of the identified specification insufficiency. In other scenarios the maneuver option of following the desired speed may still be valid if there is no noticeable crossing intention of the pedestrian.\label{fig:graph2}}

\autoref{fig:graph2} shows the adapted behavior specification in a Causal Behavior Graph. The graph contains the behavior specification for both scenarios. 
Initially, the crossing intention was inferred based on the pedestrian position being near the pedestrian crossing. 
We removed the constraint of the exact location from the fact. Thus, only the existence of a pedestrian at some position is specified as sufficient. As a result, the fact of inferring the pedestrian's crossing intention becomes more vague and requires further interpretation. Court decisions in Germany indicate that such an adaptation is legally sensible \cite{beck_how_2022} as pedestrians were in some cases allowed to cross the road without using the road markings of the pedestrian crossing. 
However, this leads to specified behavior where the ego-vehicle stops for every pedestrian that stands on the sidewalk close to a pedestrian crossing.
For further iterations of the behavior specification, 
additional operational conditions will have to be considered.
This requires, for example, to include the pedestrian's posture \cite{schneemann_context-based_2016} in the scenario description 
and will lead to adaptations in the facts and rules specified for our example.

As a result, this evaluation of our example behavior specification shows that the specified target behavior needs to be assessed regarding its correctness in addition to a formal verification of its consistency.
The evaluation also shows that the explicit documentation of assumptions in the interpretation of the traffic rules facilitates the adaptation of facts and rules. 
This consequence underpins our critique of approaches that aim for a direct formalization of traffic rules, as such approaches do not represent the context-based assumptions in a behavior specification.

For example, in comparison with Maierhofer~\emph{et~al.}~\cite{maierhofer_formalization_2020} our approach does not only provide a mapping between the legal source of a formalized rule but also documents the scenario-specific assumptions that are part of the interpretation of a traffic rule as part of the legal analysis.
Meyer~\emph{et~al.}~\cite{meyer_scenario-_2022} on the other hand establish requirements traceability at a feature level. However, with their approach the insufficiency in the interpretation of the assumed pedestrian position would be difficult to identify as the considered operating conditions are summarized in the model with the specified maneuver in a scenario.

\section{Discussion of the Semantic Norm Behavior Analysis}
\label{sec:eval}
In this section, we verify the Semantic Norm Behavior Analysis based on the requirements derived in \autoref{sec:req}, and discuss limitations of the presented approach.

\subsection{Requirements-based Verification}
\label{subsec:reqeval}
The Semantic Norm Behavior Analysis entails two fundamental parts. Firstly, knowledge sources that contain potentially conflicting and vague stakeholder needs towards the behavior of an automated vehicle are selected and analyzed. Secondly, the stakeholder needs are consolidated and formalized in a scenario-specific context in order to represent stakeholder requirements towards target behavior. As a result, the Semantic Norm Behavior Analysis satisfies the requirement to account for the needs of relevant stakeholders in the specified behavior (req. \ref{req:need}).

Requirement \ref{req:adapt} specifies that a behavior specification shall be adaptable to shifts in stakeholder needs and implementation limitations. Both were not assessed as part of our limited case study. While the Semantic Norm Behavior Analysis provides a methodology to analyze behavioral needs and formalize facts and rules, interpretations and solutions for conflicting needs can only be formulated by domain-experts (e.g. legal experts). In our case, study example, we provided an example of how ambiguous legal guidelines can be interpreted and consolidated in the context of a pedestrian crossing. For a larger scenario catalog, our assumptions need to be reexamined, and further knowledge sources need to be considered to assess the adaptability of the proposed behavior specification to shifts in stakeholder needs.

Consolidation of stakeholder requirements results in specified facts and rules. 
Facts and rules build a set of formal axioms that represent the specified behavior. Automated consistency evaluation by a reasoner supports the iterative refinement of specified behavior in the considered scenarios. Therefore, the ontology-based approach of the Semantic Norm Behavior Analysis satisfies the requirement of a consistent behavior specification in considered scenarios (req. \ref{req:consist}).

The ontology-based approach proposed in this article relies on two ontologies. The behavior specification ontology specifies facts and rules that represent causal relations between operational conditions and maneuver options. The operational domain ontology specifies context elements and their relations in the target operational domain of an automated vehicle. In the Semantic Norm Behavior Analysis, facts and rules are formalized that have explicit connections to relevant context elements in the considered scenarios. As a result, the specified behavior refers to a consistent representation of the operational environment (req. \ref{req:operat}). Additionally, the formal rules ensure that facts and rules are only applied to scenarios, in which context elements are instantiated in the operational domain ontology (req. \ref{req:scen}).

Requirement \ref{req:capab} regarding the identification of system capabilities is not covered by the discussed approach and hence considered in future work. In the aviation domain, Preece~\emph{et~al.}~\cite{preece_matching_2008} and Smirnov~\emph{et~al.}~\cite{smirnov_use_2017} published promising results on the use of ontologies to formally connect a behavior specification (or \emph{mission tasks}) with necessary capabilities.

A behavior specification that is generated with the Semantic Norm Behavior Analysis is explicitly based on stakeholder needs. As a result, behavioral requirements specify the behavior of an automated vehicle in its operational context. These requirements do not presume specific system functions or technical components (req. \ref{req:tech}). However, a vehicle's behavior will be evaluated with respect to the specified behavior \cite[Clause~10.1]{noauthor_road_2022-1}. Therefore, a refinement of the behavior specification that accounts for implementation limitations is necessary (req. \ref{req:adapt}). We did not analyze how such limitations can be accounted for in the specified behavior, as the focus of this article is to propose a methodology for a behavior specification that is explicitly based on stakeholder needs.

According to requirement \ref{req:assess} a behavior specification shall provide criteria and requirements for the assessment of multiple maneuver options. Our case study example is limited to a qualitative assessment of the compliance between specified behavior and the considered knowledge sources (i.e. the German Road Traffic Regulation). Additionally, a quantitative assessment of the specified behavior is necessary -- especially in the context of assessing the absence of unreasonable risk in a SOTIF context. While we did not analyze, how the proposed behavior specification can be assessed quantitatively, we did provide an example of a qualitative identification of a specification insufficiency in \autoref{subsec:eval}.

Requirement \ref{req:abstr} specifies that a behavior specification should not presume one level of abstraction, but that the specification should support a refinement process. The Semantic Norm Behavior Analysis relies on a scenario-based approach. We confined ourselves to the application of the Semantic Norm Behavior Analysis to functional scenarios. In the scope of our case study, we showed a possible application of the Semantic Norm Behavior Analysis at this level of abstraction. However, it is an open challenge to apply the Semantic Norm Behavior Analysis to logical and concrete scenarios, which include concrete parameters of the operational environment.

\subsection{Limitations}
\label{subsec:crit}
The first notable limitation of our application example of the Semantic Norm Behavior Analysis is that we did only consider parts of the German Road Traffic Regulation. This scope does not provide the basis for a complete set of behavioral requirements that are compliant with the entire traffic code. Additionally, an analysis that is limited to a regulatory perspective does not include all relevant stakeholder needs. 

Second, the Semantic Norm Behavior Analysis is applied in an example case study and formulated from a German perspective, as we confined ourselves mostly to the German Road Traffic Regulation\footnote{This results in particular from the fact that behavioral law is not harmonized within the European Union. This means that each member state has its own legal framework, which has a direct impact on the behavior specification, as behavioral law can differ considerably.}.
As legal methodology can diverge depending on the legal system, the scope of our example is limited to the German legal context. For example, Yu~\emph{et~al.}~\cite{yu_online_2024} propose an article by article formalization of traffic regulations in the Law of the People’s Republic of China on Road Traffic Safety. Even in Germany, in addition to established methods, legal literature and case law are also of relevance, which need to be considered.

Third, our example is limited to two scenarios which are sufficient to illustrate the application of the method 
and only allow limited conclusions regarding generalizability of the specified facts and rules and thus scalability of the Semantic Norm Behavior Analysis. For similar scenarios that can be captured as part of a use case (as defined in ISO~21448~\cite[Clause~3.32]{noauthor_road_2022-1}) similar operating conditions and thus behavioral requirements apply. However, questions regarding the scalability of any approach following a scenario-based paradigm can be raised, as generalizability and traceability are trade-off decisions. 
Favarò~\emph{et~al.}~\cite{favaro_building_2023} provide a simplified yet helpful perspective on this issue, with a focus on acceptance criteria. They define a matrix of event-level reasoning and aggregate-level reasoning, where an appropriate reasoning strategy requires both. Translated to behavior specification, the scenario-specific documentation of assumptions in the specification enables the identification of specification insufficiencies including their triggering conditions as required by ISO~21448. In how far specification insufficiencies can be generalized from an event level to an aggregate level for the entire operational domain needs to be investigated in future research.

Furthermore, the scalability of the technical implementation in this article is not investigated. The presented implementation in that regard is limited in multiple ways. As instances in the ontology are currently modeled manually per scenario, modeling errors by the developers cannot be uncovered by the logic-based reasoner. To that end, manual validation of the resulting behavior specification is necessary. Both the manual modeling and model validation can cause efficiency issues regarding modeling effort. However, modeling effort does not only apply to our approach, specifically, but to any model-based systems engineering approach. Further investigation in an actual development context is necessary to draw conclusions in how far the Semantic Norm Behavior Analysis differs in its efficiency from other model-based approaches. In contrast, the reasoning process for the formal specification is already automated. This causes a different issue with respect to filtering knowledge that is part of the ontology. Currently, the reasoner infers all statements that can be derived based on first-order predicate logic and the logic programming rules. This leads to inferred statements that are in some instances redundant and need to be identified as such by the developer, who evaluates the result of the reasoning process.

Finally, we argued that an ontology-based approach formally represents a common language among stakeholders. However, the specific ontologies that are used in our application examples are subject to ontological uncertainties \cite{spiegelhalter2017} and are limited to the scope of the considered scenarios. To use these ontologies at a larger scale, relevant stakeholders need to be included in the definition of the modeled concepts.

\section{Conclusion and Future Work}
Current safety standards such as ISO~21448~\cite{noauthor_road_2022-1} and regulations in the European Union \cite{noauthor_commission_2022} indicate that an explicit specification of the intended behavior of an automated vehicle supports the assurance of its safety and compliance with traffic rules. However, neither in normative documents nor in related research, formalized approaches are proposed to support the processes leading to a behavior specification.

In this article, we identified necessary research as well as requirements (\autoref{sec:req}) that are -- in part -- implicitly raised towards a behavior specification in related work (\autoref{sec:relw}). Additionally, we proposed terminology (\autoref{sec:term}) to describe some of the key concepts that -- based on our analysis of the related work -- have not yet been adequately defined.

To address some of the identified challenges in this article, we introduced the Semantic Norm Behavior Analysis as an ontology-based approach to support a traceable behavior specification in automated driving (\autoref{sec:snba}). More specifically, we focused on the explicit and formal documentation of assumptions and decisions under specified conditions that are part of a behavior specification.

We showed how the Semantic Norm Behavior Analysis provides a means to explicitly specify the intended behavior of an automated vehicle in an example scenario (\autoref{sec:examp}). In this example, we showed that the proposed method facilitates the consolidation of behavioral requirements based on parts of the German Road Traffic Regulation. 
In this context, we introduced facts and rules as key concepts in the behavior specification to capture causalities between the operational environment of an automated vehicle and the specified maneuver options.
We evaluated the results of the example behavior specification. The key finding in the evaluation was that the ontology-based behavior specification supported the identification of a specification insufficiency under slightly varied conditions in the example scenario. 

Finally, based on the evaluation of the proposed approach, we can formulate challenges for future work.
First, while the Semantic Norm Behavior Analysis provides an approach to generate a behavior specification, the integration into a larger system and safety engineering process was not shown in this article. To contribute to SOTIF activities beyond the identification of specification insufficiencies, such an integration is necessary. 
To the related end of a consistent knowledge representation of the Operational Design Domain throughout the development process, Stierand~\emph{et~al.}~\cite{stierand_using_2024} elaborate on potential benefits utilizing ontologies.

Second, this article is limited by the considered level of abstraction of specified behavior. We showed an ontology-based approach to explicitly specify behavior in functional scenarios. However, for more detailed SOTIF analyses, behavior needs to be specified in a logical scenario including parameterized operating conditions \cite{zhang_odd_2024}. Such analyses are, for example, sensitive to hardware or software design and can uncover specific triggering conditions.

Third, we confined ourselves to one specific set of sources for the example behavior specification. Mainly, we used parts of the German Road Traffic Regulation. As a result, the kind of stakeholder needs that are represented in the behavior specification is limited. Further sources that can lead to a more adequate representation of stakeholder needs (in Germany) are, for example, court cases, and non-legal sources such as recommendations by the German ethics committee.

Fourth, the examples in this article are limited to a German legal perspective and are based on a minimal set of regulations. Other legal contexts are not analyzed in this article. Thus, the results in this article cannot be simply transferred. To generate a behavior specification at a larger scale, the sensitivity of the specified behavior to differences, for example, in the prevailing law, but also the driving style need to be analyzed.

Future work could investigate how our formalization in description logic could be complemented by approaches relying on, for example, temporal logic (e.g. \cite{maierhofer_formalization_2020}). Such approaches would be especially relevant for the specification in more detailed scenarios.

Additionally, with respect to the challenge of manual modeling effort and potential errors we see high potential in partial automation of the specification process, for example, by generating ontology instances based on a specification in the domain-specific modeling language stiEF \cite{bock_advantageous_2019}. Such a tailored language has the benefit of easier accessibility for developers to adopt and is already integrated into industry tools. As a result, the modeling effort in the ABox of the ontology could be reduced.

A field of research that we did not address as part of our work is the technical realization of the specified behavior. Data-driven approaches using machine learning are an important improvement of runtime algorithms in automated vehicles. However, we focused on how to support developers with means to document their assumptions during development. A promising work on how the gap in knowledge representation at design time and runtime can be bridged is presented by Patrikar~\emph{et~al.}~\cite{patrikar_rulefuser_2024}. With the \emph{RuleFuser} framework the authors show, that a planner can benefit in out-of-distribution scenarios by combining a rule-based and a machine learning approach. Thus, we see potential in combination of our approach with data-driven framework such as presented by Patrikar~\emph{et~al.}~\cite{patrikar_rulefuser_2024}.

\section*{Acknowledgment}
We thank our industry colleagues Christian Lalitsch-Schneider (ZF Friedrichshafen AG), Hans Nikolaus Beck (Daimler Truck AG, formerly Robert Bosch GmbH), and Dr.\,Christoph Höhmann (Mercedes-Benz AG) for contributing to the discussions regarding this work. Special thanks go to the students Diyar Polat, and Christian Kühr who supported this work with their work at the Institute of Control Engineering (Technische Universität Braunschweig). 
Last but not least we would like to sincerely thank the reviewers for their valuable suggestions that helped to improve this manuscript.

\bibliographystyle{IEEEtran}
\bibliography{snba}

\begin{IEEEbiography}[{\includegraphics[width=1in,clip,keepaspectratio]{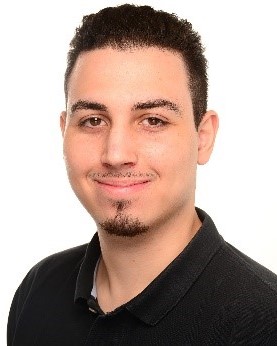}}]{Nayel Fabian Salem} received the B.Sc. degree in mechanical engineering from Technische Universit\"at Berlin (2018), Berlin, Germany, and the M.Sc. degree in electronic systems engineering from the Technische Universität Braunschweig (2020), Braunschweig, Germany.
Since 2021, he has been a Research Associate pursuing his PhD at the Institute of Control Engineering, Technische Universit\"at Braunschweig. His main research interests include safety assurance of automated vehicles, focusing on safety issues regarding behavior specification.
\end{IEEEbiography}

\begin{IEEEbiography}[{\includegraphics[width=1in,clip,keepaspectratio]{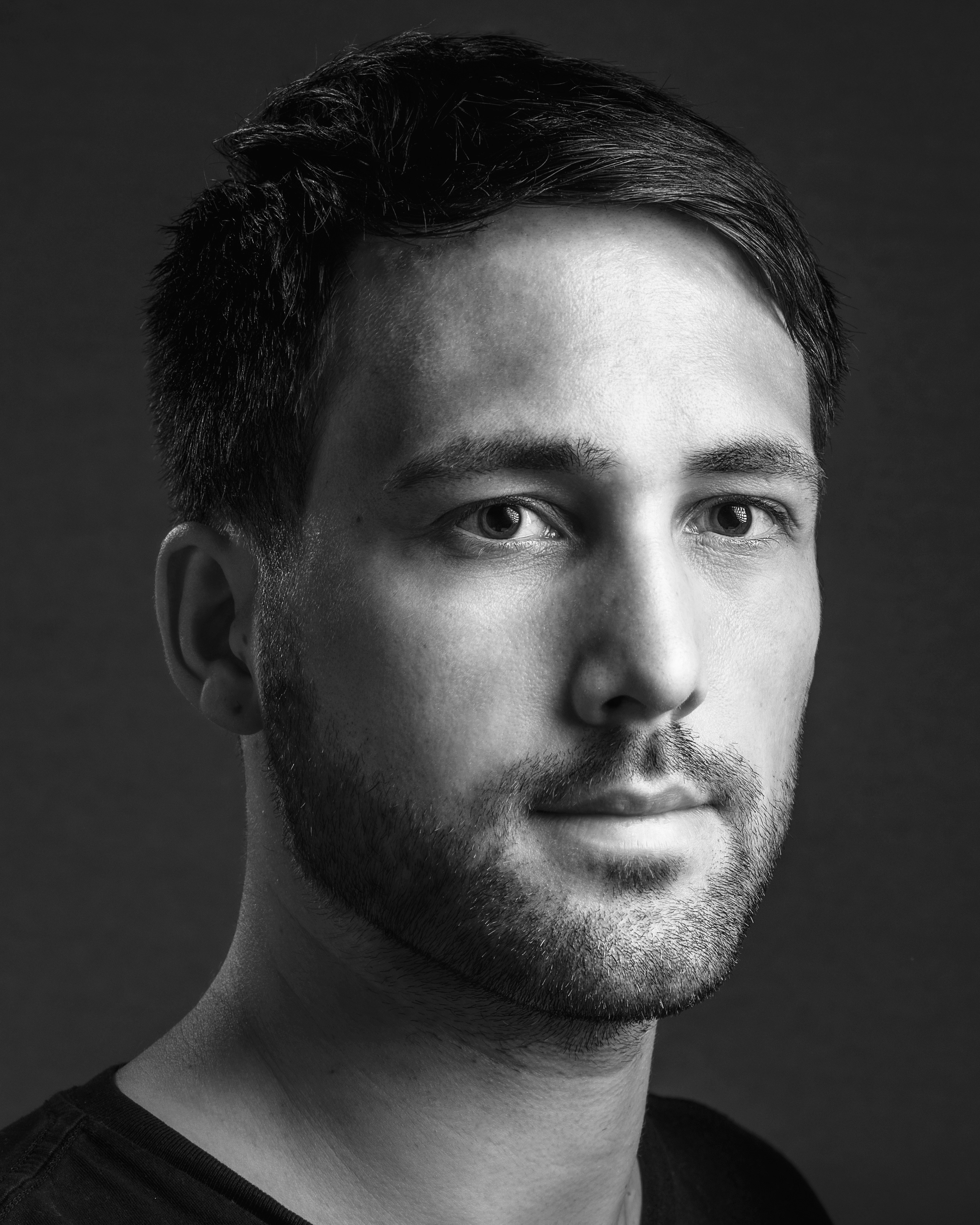}}]{Marcus Nolte} works as a Research Associate at the Institute of Control Engineering at TU Braunschweig since 2014 and is currently pursuing his PhD. He received his Bachelor and Master of Science in electrical engineering from TU Braunschweig in 2011 and 2014. 
His main research interest is self- and risk-aware and motion planning for automated vehicles.
\end{IEEEbiography}

\begin{IEEEbiography}[{\includegraphics[width=1in,clip,keepaspectratio]{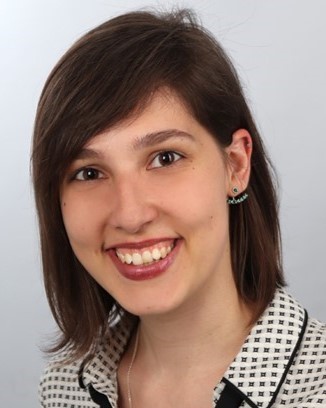}}]{Veronica Haber} has been working as a Systems Engineering Consultant at PROSTEP since 2018, contributing to various projects in the automotive industry. She received her M.Sc  degree in Systems Engineering from the Hochschule für angewandte Wissenschaften München in 2019 and her B.Sc degree in Management \& Technology from the Technische Universität München in 2016. She aims to enable the integration of sustainable mobility concepts for future generations.
\end{IEEEbiography}

\begin{IEEEbiography}[{\includegraphics[width=1in,clip,keepaspectratio]{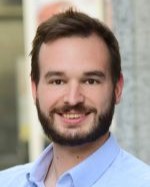}}]{Till Menzel} received the B.Sc. and M.Sc.
degrees in electrical engineering from the Technische Universität Braunschweig, Braunschweig, Germany, in 2011 and 2014, respectively, where he is currently pursuing the Ph.D. degree with
the Institute of Control Engineering. 
His main research interests include scenario-based verification and validation of automated vehicles, focusing on a systematic generation of scenarios for
simulation-based testing.
\end{IEEEbiography}

\begin{IEEEbiography}[{\includegraphics[width=1in,clip,keepaspectratio]{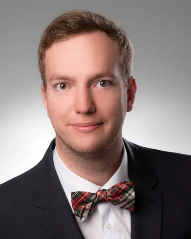}}]{Hans Steege} holds a doctorate in law from the Leibniz University of Hanover and a doctorate in economics and social sciences from the University of Stuttgart. He worked in the legal department of Continental AG, in public affairs at Volkswagen Commercial Vehicles and has been working in data protection at Cariad SE, a Volkswagen Group Company, since 2021.
He is a lecturer at the University of Stuttgart, where he teaches in the fields of autonomous driving and artificial intelligence.
He researches, teaches, writes legal opinions, and publishes on legal aspects of autonomous driving, artificial intelligence and the metaverse. One focus is on the implications of product compliance and the programming of traffic regulations. In the contribution, he exclusively reflects his personal legal opinion.
\end{IEEEbiography}

\begin{IEEEbiography}[{\includegraphics[width=1in,clip,keepaspectratio]{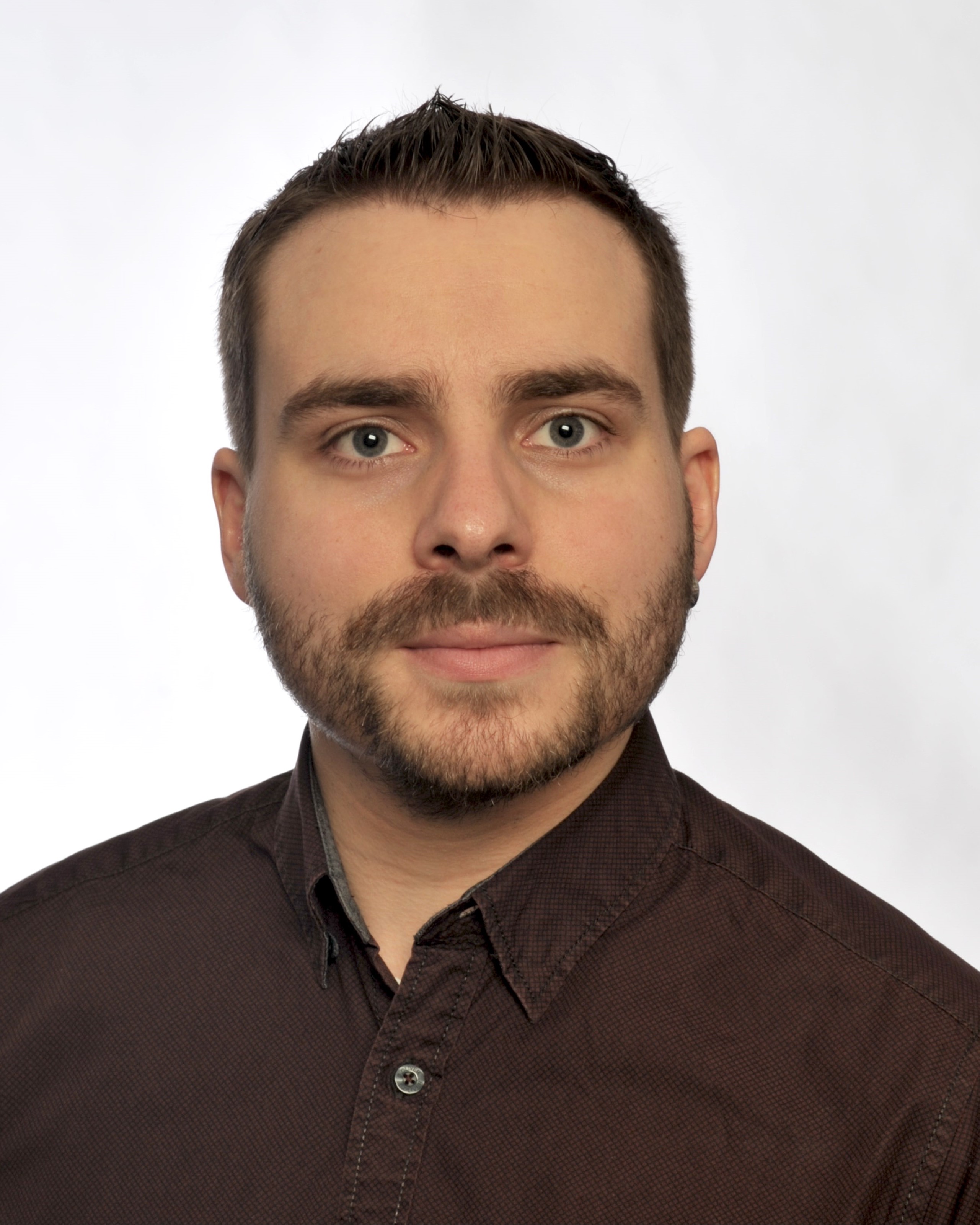}}]{Robert Graubohm} received the B.Sc. (2013) and M.Sc. (2016) degree in
industrial engineering in the field mechanical engineering from Technische Universität Braunschweig, Germany, and the M.B.A. (2015) degree
from the University of Rhode Island, Kingston, RI, USA. 
He is currently a Research Associate with the Institute of Control Engineering,
Technische Universität Braunschweig. His main research interests include development processes of automated driving functions and the safety conception
in an early design stage.
\end{IEEEbiography}

\begin{IEEEbiography}[{\includegraphics[width=1in,clip,keepaspectratio]{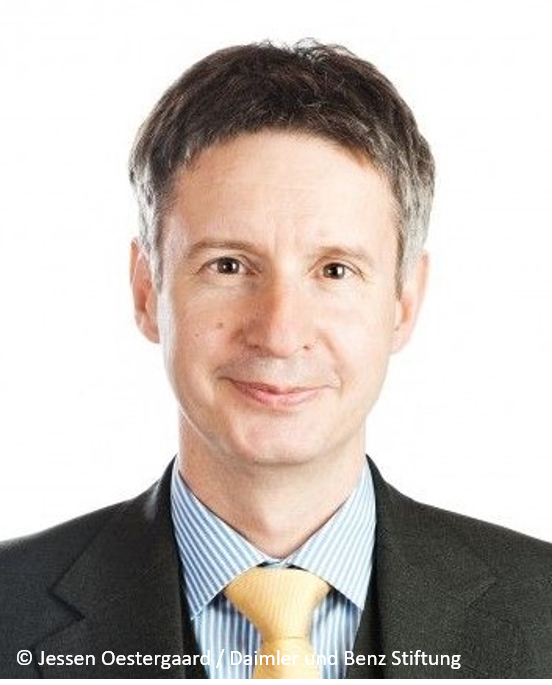}}]{Markus Maurer} received the Diploma degree in electrical engineering from the Technische Universität München, in 1993, and the PhD degree in automated driving from the Group of Prof. E. D. Dickmanns, Universität der Bundeswehr München, in 2000. 
From 1999 to 2007, he was a Project Manager and the Head of the Development Department of Driver Assistance Systems, Audi. 
Since 2007, he has been a Full Professor of automotive electronics systems with the Institute of Control Engineering, Technische Universität Braunschweig.
His research interests include both functional and systemic aspects of automated road vehicles.
\end{IEEEbiography}

\EOD

\end{document}